\documentclass[longnamesfirst,apj]{emulateapj}
\usepackage{apjfonts,natbib}
\tighten

\def\xflux{~erg~cm$^{-2}$~s$^{-1}$}

\received{2004 May 18}
\accepted{2004 July 28}

\slugcomment{The Astronomical Journal, in press}
\shortauthors{BAUER ET AL.}
\shorttitle{CHANDRA DEEP FIELD X-RAY NUMBER COUNTS}

\begin{document}

\title{The Fall of AGN and the Rise of Star-Forming Galaxies:
A Close Look at the {\it Chandra} Deep Field X-ray Number 
Counts}

\author{ F.~E.~Bauer,\altaffilmark{1, 2}
  D.~M.~Alexander,\altaffilmark{1} W.~N.~Brandt,\altaffilmark{2}
  D.~P.~Schneider,\altaffilmark{2}
  E.~Treister,\altaffilmark{3, 4}
  A.~E.~Hornschemeier,\altaffilmark{5, 6} and
  G.~P.~Garmire\altaffilmark{2} }

\altaffiltext{1}{Institute of Astronomy, University of Cambridge, 
Madingley Rd., Cambridge, CB3 0HA, UK}
\altaffiltext{2}{Department of Astronomy \& Astrophysics, 525 Davey Lab, 
The Pennsylvania State University, University Park, PA 16802, USA}
\altaffiltext{3}{Yale Center for Astronomy \& Astrophysics, Yale
  University, P.O. Box 208121, New Haven, CT 06520}
\altaffiltext{4}{Departamento de Astronom\'{\i}a, Universidad de
  Chile, Casilla 36-D, Santiago, Chile}
\altaffiltext{5}{Department of Physics and Astronomy, Johns Hopkins
University, 3400 North Charles Street, Baltimore, MD 21218}
\altaffiltext{6}{Chandra Fellow}

\begin{abstract}
  We investigate the X-ray number counts in the 1--2~Ms {\it Chandra}
  Deep Fields (CDFs) to determine the contributions of faint X-ray
  source populations to the extragalactic X-ray background
  (XRB).
  X-ray sources were separated into Active Galactic Nuclei (AGN),
  star-forming galaxies, and Galactic stars based primarily on
  X-ray-to-optical flux ratios, optical spectral classifications,
  X-ray spectra, and intrinsic X-ray luminosities. 
  Number-count slopes and normalizations below $2\times10^{-15}$
  erg~cm$^{-2}$~s$^{-1}$ were calculated in each band for all source
  types assuming a single power-law model.
  We find that AGN continue to dominate the number counts in the
  0.5--2.0~keV and 2--8~keV bands. At flux limits of $\approx
  2.5\times 10^{-17}$~erg~cm$^{-2}$~s$^{-1}$ (0.5--2.0~keV) and
  $\approx 1.4\times 10^{-16}$~erg~cm$^{-2}$~s$^{-1}$ (2--8~keV), the
  overall AGN source densities are 7166$^{+304}_{-292}$
  sources~deg$^{-2}$ and 4558$^{+216}_{-207}$ sources~deg$^{-2}$,
  respectively; these are factors of $\sim10$--20 higher than found in
  the deepest optical spectroscopic surveys.
  While still a minority, the number counts of star-forming galaxies
  climb steeply such that they eventually achieve source densities of
  1727$^{+187}_{-169}$ sources~deg$^{-2}$ (0.5--2.0~keV) and
  711$^{+270}_{-202}$ sources~deg$^{-2}$ (2-8~keV) at the CDF flux
  limits. The number of star-forming galaxies will likely overtake the
  number of AGN at $\sim1\times10^{-17}$~erg~cm$^{-2}$~s$^{-1}$
  (0.5--2.0~keV) and dominate the overall number counts thereafter.
  Adopting XRB flux densities of
  ($7.52\pm0.35)\times10^{-12}$~erg~cm$^{-2}$~s$^{-1}$~deg$^{-2}$
  (0.5--2.0~keV) and
  ($2.24\pm0.11)\times10^{-11}$~erg~cm$^{-2}$~s$^{-1}$~deg$^{-2}$
  (2--8~keV), the CDFs resolve a total of 89.5$^{+5.9}_{-5.7}$\% and
  86.9$^{+6.6}_{-6.3}$\% of the extragalactic 0.5--2.0~keV and
  2--8~keV XRBs, respectively.
  AGN as a whole contribute $\approx$83\% and $\approx$95\% to the
  these resolved XRB fractions, respectively, while star-forming
  galaxies comprise only $\approx$3\% and $\approx$2\%, respectively,
  and Galactic stars comprise the remainder.
  Extrapolation of the number-count slopes can easily account for the
  entire 0.5--2.0~keV and 2--8~keV XRBs to within statistical errors.
  We additionally examine the X-ray number counts as functions of
  intrinsic X-ray luminosity and absorption, finding that sources with
  $L_{\rm 0.5-8~keV}>10^{43.5}$~erg~s$^{-1}$ and $N_{\rm
    H}<10^{22}$~cm$^{-2}$ are the dominant contributors to the
  0.5--2.0~keV XRB flux density, while sources with $L_{\rm
    0.5-8~keV}=10^{42.5}$--$10^{44.5}$~erg~s$^{-1}$ and a broad range
  of absorption column densities primarily contribute to the 2--8~keV
  XRB flux density. This trend suggests that even less intrinsically
  luminous, more highly obscured AGN may dominate the number counts at
  higher energies where the XRB intensity peaks.
  Finally, we revisit the reported differences between the CDF-North
  and CDF-South number counts, finding that the two fields are
  consistent with each other except for 2--8~keV detected sources
  below $F_{\rm 2-8~keV}\approx1\times10^{-15}$~erg~cm$^{-2}$~s$^{-1}$,
  where deviations gradually increase to $\approx3.9\sigma$.
\end{abstract}


\section{Introduction}\label{sec:intro}

Deep {\it Chandra} and {\it XMM-Newton} observations have now resolved
the vast majority of the X-ray background (XRB) below $\approx$8~keV
\citetext{e.g., \citealp{Cowie2002}, hereafter C02;
  \citealp{Moretti2003}, hereafter M03; \citealp{Worsley2004}}, with
much of the remaining uncertainty in the resolved fraction attributed
to deviations in the absolute value of the XRB itself due to
large-scale structure \citep[e.g.,][]{Gilli2003b,Yang2003} and
significant instrumental cross-calibration uncertainties
\citep[e.g.,][and references therein]{DeLuca2004}.  The number counts
in both the 0.5--2.0~keV (soft) and 2--8~keV (hard) bands can be
fitted with characteristic double power-law shapes with breaks around
$10^{-15}$--$10^{-14}$~erg~cm$^{-2}$~s$^{-1}$ (e.g., CO2; M03).
Intensive optical follow-up campaigns have shown that extragalactic
sources at bright X-ray fluxes ($\ga10^{-13}$--$10^{-14}$ \xflux) are
generally unobscured or mildly obscured active galactic nuclei
\citep[AGN; e.g.,][]{Bade1998, Schmidt1998, Akiyama2003}, while below
this level several other populations emerge such as obscured AGN
\citep[e.g.,][]{Alexander2001, Tozzi2001, Barger2002, Mainieri2002}
and starburst and quiescent galaxies \citep[e.g.,][]{Giacconi2001,
  Alexander2002b, Bauer2002b, Hornschemeier2003a}.

This paper builds upon the overall number-count results of C02 and
M03, which were both based on the 1~Ms {\it Chandra} Deep Field (CDF)
datasets, and extends the quiescent galaxy number-count results
presented by \citet{Hornschemeier2003a} by investigating how various
source populations contribute to the XRB using the multi-wavelength
datasets of the 2~Ms CDF-North (\hbox{CDF-N}) and 1~Ms CDF-South
(\hbox{CDF-S}). In addition to deep X-ray observations, these legacy
fields have deep {\it HST}, radio, and ground-based optical imaging,
as well as several thousand spectroscopic redshifts, allowing
classification of different source types.  We describe our X-ray
sample in $\S$\ref{sec:data}, while our method for estimating
incompleteness and bias is outlined in $\S$\ref{sec:simulations}.
Finally, we present the selection and classification of sources and
number-count results in $\S$\ref{sec:number_counts}. Throughout this
paper, we adopt $H_{0}=70$~km~s$^{-1}$~Mpc$^{-1}$, $\Omega_{\rm
  M}=0.3$, and $\Omega_{\Lambda}=0.7$ \citep{Spergel2003}. Unless
explicitly stated otherwise, quoted errors are for a $1\sigma$ (68\%)
confidence level.

\section{X-ray Sample}\label{sec:data}

\begin{figure}
\vspace{-0.1in}
\centerline{
\includegraphics[width=9.0cm]{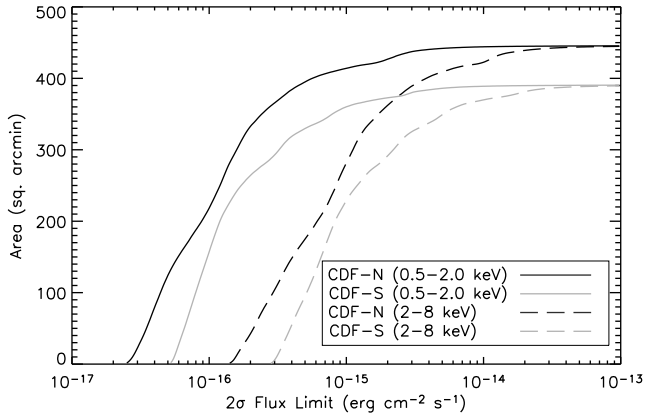}
}
\vspace{-0.1cm} 
\figcaption[Fig.1]{
  The \hbox{CDF-N} (black curves) and \hbox{CDF-S} (grey curves) solid angle
  coverage for a 2$\sigma$ flux limit. Solid and dashed curves
  represent solid angle coverage in the 0.5--2.0~keV and 2--8~keV
  bands, respectively.
\label{fig:area_vs_flux}}
\vspace{-0.3cm} 
\end{figure} 

Our sample is derived from the catalogs of \citet[][hereafter
A03]{Alexander2003b}, which consist of 503 X-ray sources in the 2~Ms
\hbox{CDF-N} and 326 X-ray sources in the 1~Ms \hbox{CDF-S}. We have chosen to
combine these two samples to achieve a more accurate census of the
X-ray source population. We recognize that the number counts of these
two fields have been reported to differ by $\sim30$\% at faint X-ray
fluxes (e.g., C02; M03), and we briefly investigate such differences in
$\S$\ref{sec:number_counts}. The on-axis sensitivity limits for the
\hbox{CDF-N} and \hbox{CDF-S} are $\approx 2.5\times
10^{-17}$~erg~cm$^{-2}$~s$^{-1}$ and $\approx
5.2\times10^{-17}$~erg~cm$^{-2}$~s$^{-1}$ (soft) and $\approx
1.4\times 10^{-16}$~erg~cm$^{-2}$~s$^{-1}$ and $\approx
2.8\times10^{-16}$~erg~cm$^{-2}$~s$^{-1}$ (hard), respectively.  X-ray
source fluxes are taken directly from these catalogs and include
corrections for vignetting, individual spectral slopes when known
(otherwise $\Gamma=1.4$ is assumed), and contamination of the ACIS blocking
filters. 

We have additionally corrected the fluxes for the intervening Galactic
column densities.\footnote{The Galactic column~densities toward the
  \hbox{CDF-N} and \hbox{CDF-S} are $(1.3\pm0.4)\times10^{20}$
  cm$^{-2}$ and $(8.8\pm4.0)\times10^{19}$ cm$^{-2}$, respectively
  \citep{Lockman2004,Stark1992}.}  For our number-count estimates in
each band, we have imposed a flux cutoff to the above sample based on
the 2$\sigma$ limiting flux maps for the \hbox{CDF-N} and
\hbox{CDF-S}, which were constructed following the prescription in
$\S$4.2 of A03. This significance level empirically matches our
limiting sensitivity across the field and is adopted to remove a small
number of sources with large and highly uncertain completeness and
flux bias corrections (i.e., $\la10$\% completeness; see
$\S$\ref{sec:simulations}). Figure~\ref{fig:area_vs_flux} presents the
sensitivity and sky coverage of the CDFs. Twenty soft-band and 10
hard-band sources were rejected because they were below this flux
threshold. In total, we used 724 sources detected in the soft band and
520 sources detected in the hard band to estimate the X-ray number
counts, with sky coverages ranging from 0.232 deg$^{2}$ to 0.004
deg$^{2}$ depending on X-ray flux. Note that the CDF sample presented
here consists only of point sources and does not include any obvious
contribution from X-ray clusters or groups \citep[e.g.,
][]{Bauer2002a}. Thus our total X-ray number counts and resolved XRB
fraction estimates will likely underestimate the real quantities by a
few percent \citep[for estimates of this small contribution see,
e.g.,][]{Rosati2002}.

\section{Chandra Simulations}\label{sec:simulations}

\begin{figure*}
\vspace{-0.1in}
\centerline{
\includegraphics[width=9.0cm]{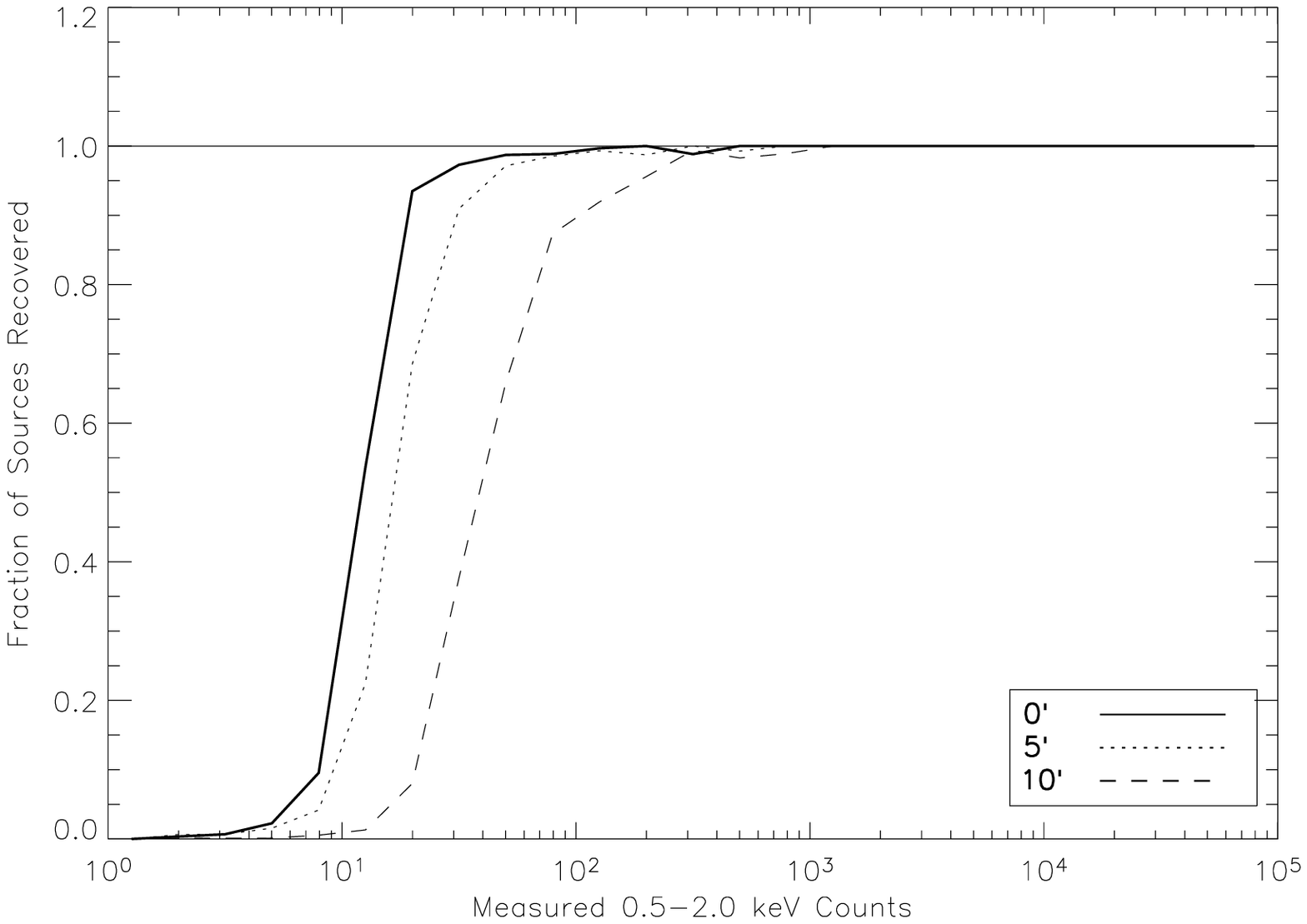}\hfill
\includegraphics[width=9.0cm]{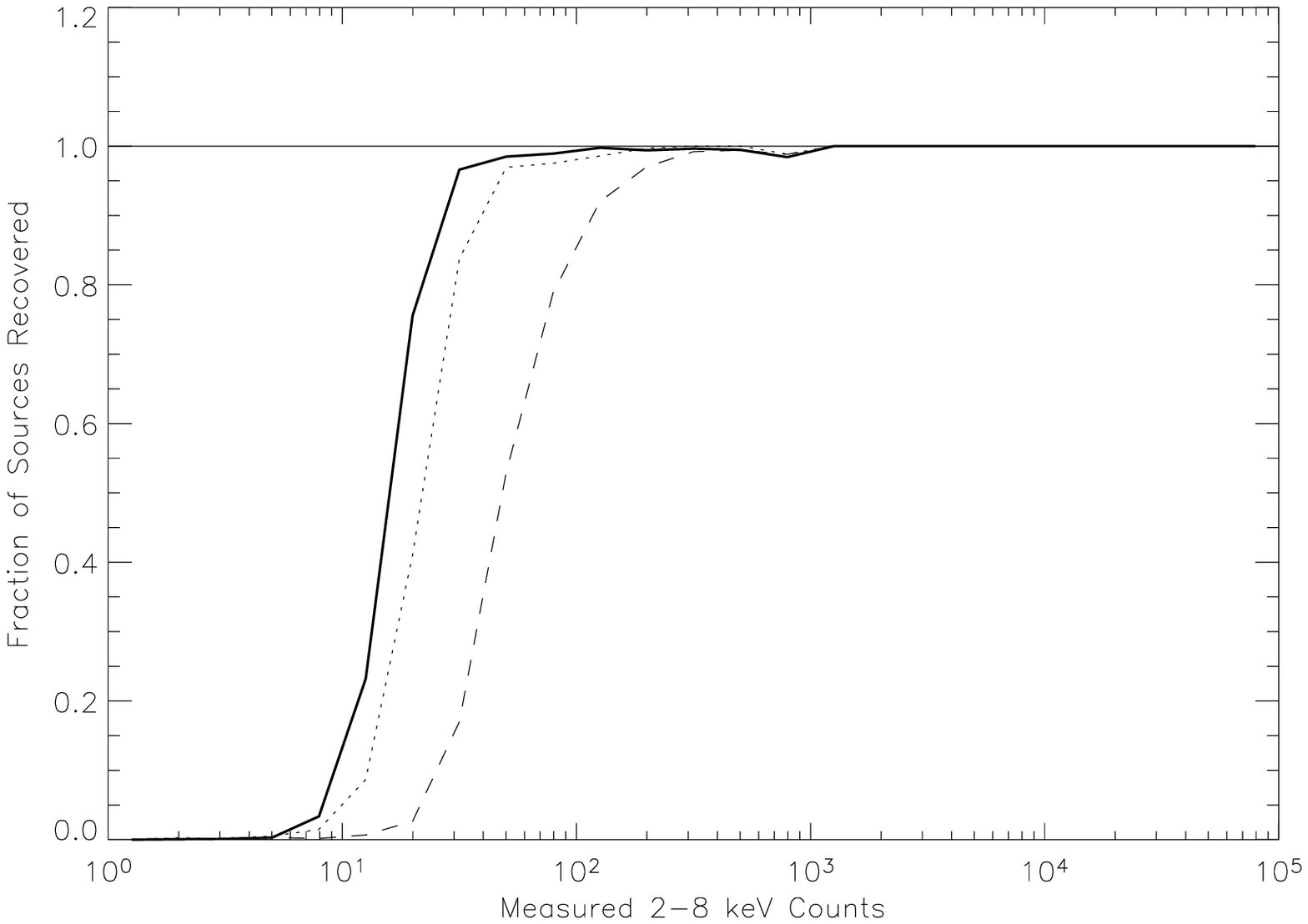}
}
\centerline{
\includegraphics[width=9.0cm]{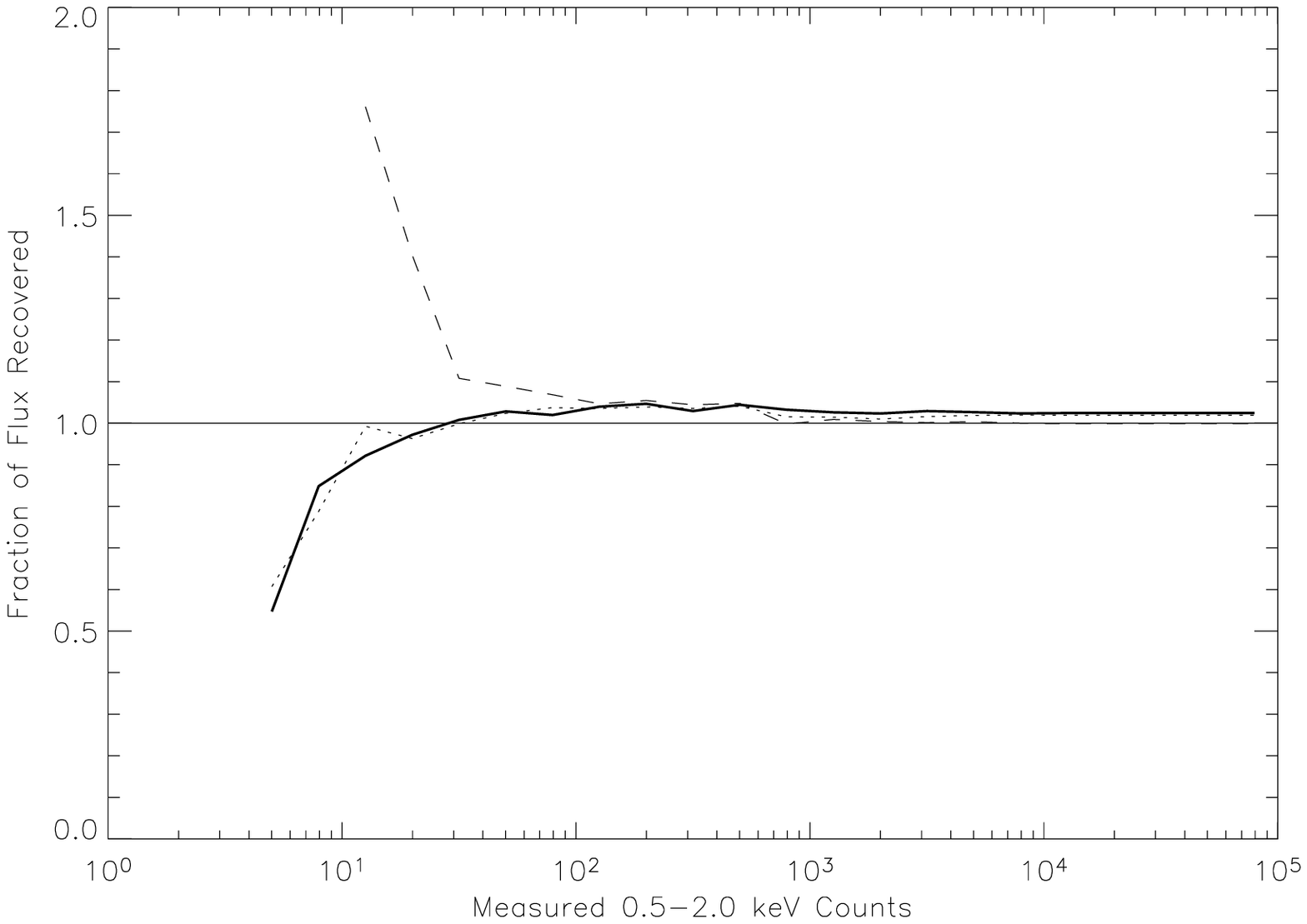}\hfill
\includegraphics[width=9.0cm]{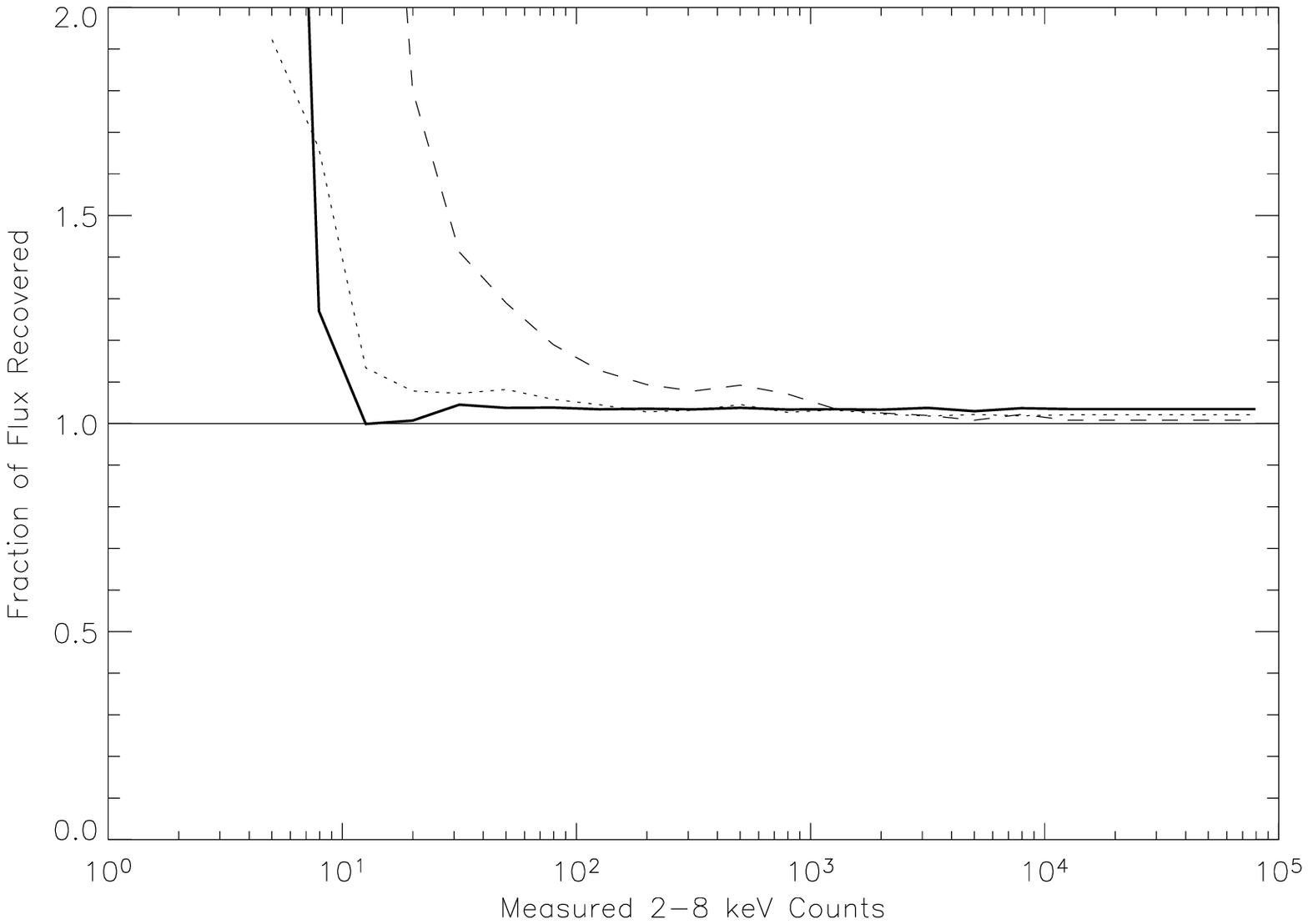}
}
\vspace{-0.1cm} 
\figcaption[Fig.2]{
  The simulated completeness ({\it top}) and flux recovery correction
  ({\it bottom}) functions for three representative positions in the
  \hbox{CDF-N} field as a function of measured counts. The solid, dotted, and
  dashed curves denote median values at off-axis angles of
  0$\arcmin$, 5$\arcmin$, and 10$\arcmin$, respectively, using all
  simulated sources within 2$\arcmin$ of each representative position;
  the thin solid line indicates the value of unity.  The soft- and
  hard-band corrections are shown on the {\it left} and {\it right},
  respectively. Eddington bias is only significant for the
  0.5--2.0~keV number counts in the 0$\arcmin$ and 5$\arcmin$ samples,
  where the recovered flux turns downward for low count sources.
\label{fig:bias_corrections}}
\vspace{-0.3cm} 
\end{figure*} 

To understand the effects of incompleteness and bias, we created 200
Monte Carlo simulated observations in both the soft and hard energy
bands for each of the CDFs. We began by creating template background
images for each field in the soft and hard bands following $\S$4.2 of
A03. To this end, we masked out all 829 known point-sources using
circular apertures with radii twice those of the $\approx$~90\% point
spread function (PSF) encircled-energy radii. We filled in the masked
regions for each source with a local background estimated by making a
Poisson probability distribution of counts using an annulus with inner
and outer radii of 2 and 4 times the $\approx$90\% PSF
encircled-energy radius, respectively (for further details see A03).
The resultant images include minimal contributions from detected point
sources while still providing realistic contributions from extended
sources \citep[e.g.,][]{Bauer2002a}, which will cause a slight
overestimation of the measured background close to extended sources.

To these template background images we added simulated sources at
random positions. The fluxes of these simulated sources were drawn
randomly from the total number-count models of M03 between
$10^{-18}$--$10^{-11}$~erg~cm$^{-2}$~s$^{-1}$ (soft) and
$10^{-17}$--$10^{-11}$~erg~cm$^{-2}$~s$^{-1}$ (hard), respectively.
These fluxes were converted to X-ray count rates assuming a
$\Gamma=1.4$ power law and the maximum effective area of the CDFs.
Assuming a somewhat different average X-ray spectral slope
($\Gamma=1$--2) does not dramatically change our results. We have not
attempted to account for any scatter in the distribution of $\Gamma$
and note that individual sources with significantly steeper or flatter
spectral slopes are likely to have much higher flux thresholds for
detection due to the energy dependence of {\it Chandra}'s effective
area. The number of CDF sources with deviant X-ray spectral slopes is
small, and therefore are unlikely to have a large effect.

\begin{figure*}
\vspace{-0.1in}
\centerline{
\includegraphics[width=9.0cm]{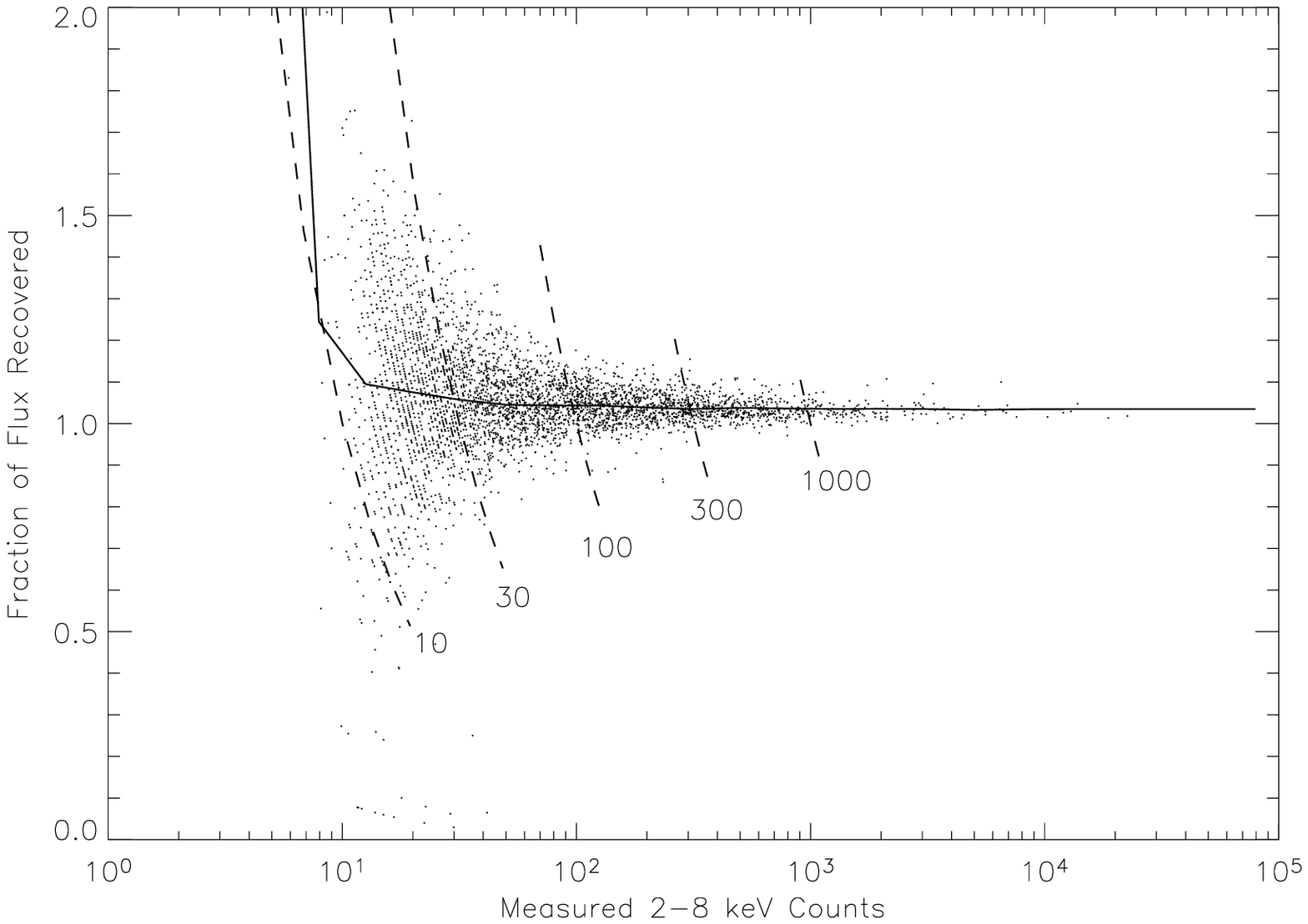}\hfill
\includegraphics[width=9.0cm]{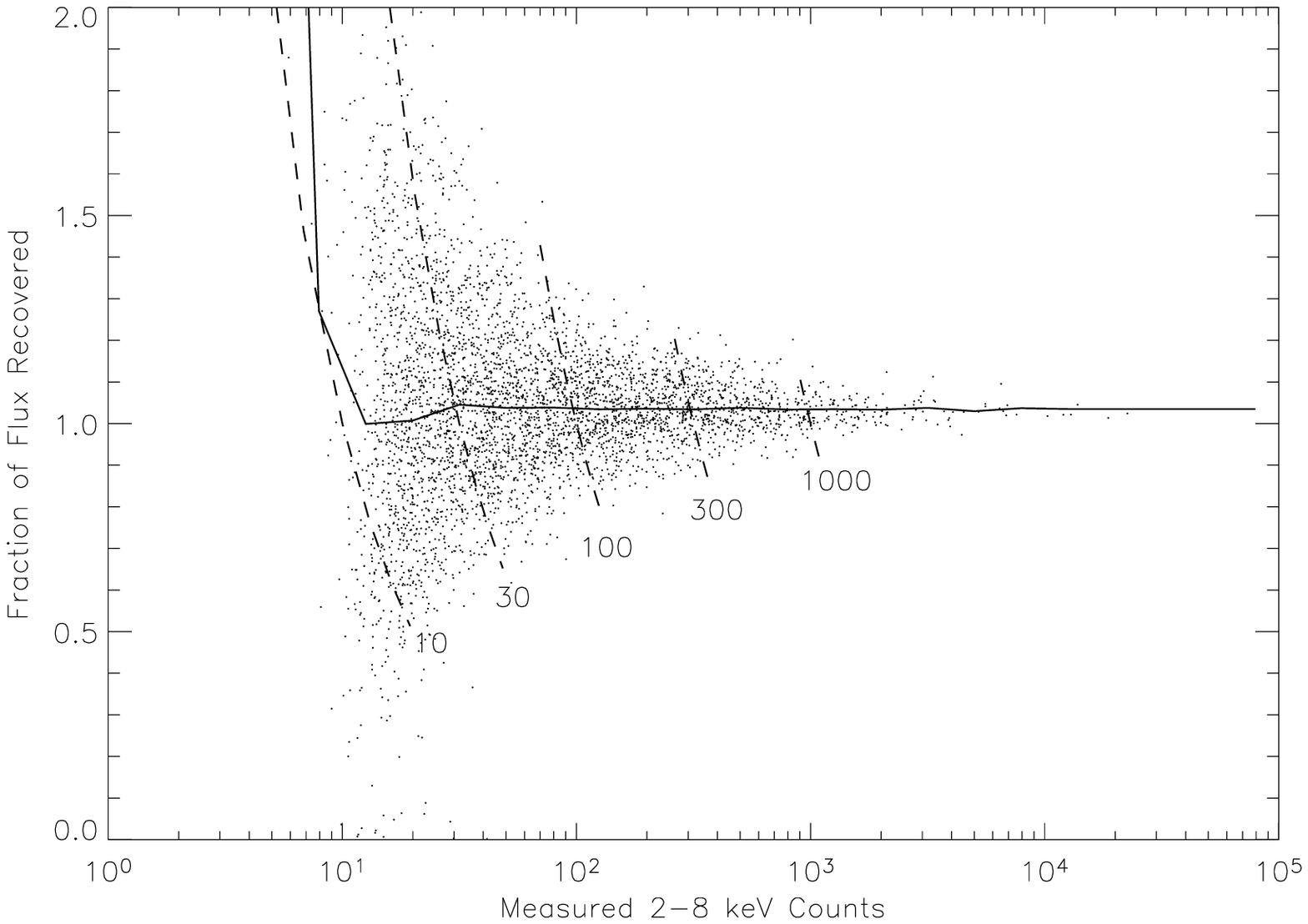}
}
\vspace{-0.1cm} 
\figcaption[Fig.3]{
  The individual flux recovery corrections for hard-band simulated
  sources within 2$\arcmin$ of the 0$\arcmin$ off-axis position shown
  in Figure~\ref{fig:bias_corrections} (these sources are represented
  by the solid curve in the lower right diagram of
  Figure~\ref{fig:bias_corrections}). The solid curve indicates the
  median flux correction, akin to those plotted in the bottom of
  Figure~\ref{fig:bias_corrections}.  The corrections are shown both
  including ({\it right}) and excluding ({\it left}) Eddington bias,
  in order to demonstrate separately the effects of our photometry
  alone and our photometry plus Eddington bias. Photometry errors
  alone skew sources diagonally in these diagrams such that
  underestimated sources have both lower counts and higher flux
  recovery corrections, while overestimated sources have both higher
  counts and lower flux recovery corrections. Thus photometry errors
  by themselves imprint a gradual upward trend in the flux corrections
  for decreasing source counts. The dashed curves indicate the sense
  of this photometry effect, showing tracks along which simulated
  sources with 10, 30, 100, 300, and 1000 counts would be scattered
  due to photometric errors. The tracks span a range of $\pm3\sigma$.
  Very faint sources with low flux corrections tend to be simulated
  sources which lie below our nominal detection threshold but are
  detected because they sit on positive background fluctuations (both
  panels) or are caused by Eddington bias (right panel only).  Adding
  Eddington bias to the simulation injects significantly more noise
  into the distribution of flux corrections and provides several
  additional faint sources with low flux corrections. As such,
  Eddington bias tends to pull the median flux correction down for
  faint sources.
\label{fig:flux_bias_explain}}
\vspace{-0.3cm} 
\end{figure*}

Exposure times for the simulated sources were derived from their
positions on the CDF exposure maps (see $\S$3.1 of A03) and used to
convert count rates to counts. To include the effects of Eddington
bias (i.e., the measured flux is higher than the actual flux due to
statistical fluctuations), the counts for a simulated source were then
redrawn from a statistical error distribution. Note that we kept track
of the counts both before and after including Eddington bias in order
to decouple the effects of our photometry from those of Eddington bias
(see Fig.~\ref{fig:flux_bias_explain}). Finally, counts for each
source were drawn randomly from a PSF probability distribution
function to simulate a real source and then added to the template
image. To mimic the complex PSF of the multi-observation CDFs, we
adopted the combined model PSF from the nearest real X-ray source in
the CDFs.  These model PSFs were produced using the IDL-based source
extraction tool ACIS\_EXTRACT \citep[for details see][]{Broos2004},
whereby the PSF for each individual {\it Chandra} observation was
calculated with the CIAO tool MKPSF using the CALDB PSF libraries,
weighted by the number of counts in each exposure, and co-added. The
nearest real source was always $\la1\arcmin$ from the simulated source
position.

Source searching and photometry of the simulated images were performed
in a manner identical to that used to produce the CDF catalogs [i.e.,
using WAVDETECT \citep{Freeman2002} and custom software; see $\S$3.2
and $\S$3.4 of A03 for details]. A completeness correction was
determined by comparing the number of simulated input sources to the
number of simulated detected sources as a function of detected counts.
Likewise, a flux recovery correction was estimated by comparing the
simulated input counts (both prior to and after including Eddington
bias) to our measured aperture-corrected counts. Since both
completeness and flux recovery vary across the CDF fields due to {\it
  Chandra}'s changing PSF and spatially dependent vignetting, we
determined the completeness and flux recovery functions within a
2$\arcmin$ radius region around each real source, averaged over the
200 simulations.  The radius value of 2$\arcmin$ was chosen as a
compromise between the maximum area over which the PSF remains
relatively constant and the minimum area needed to achieve reasonable
statistics with our 200 simulations. The above correction functions
were used to correct the measured source densities and fluxes,
respectively.  The completeness and flux recovery functions remain
close to unity above \hbox{$\sim50$--100}~counts.  Below this point,
{\it Chandra}'s varying PSF size and spatially dependent vignetting
begin to affect source detection and photometry.
Figure~\ref{fig:bias_corrections} demonstrates how the completeness
and flux recovery functions behave for three separate positions in the
\hbox{CDF-N} field in both the soft and hard bands; similar functions were
obtained for the \hbox{CDF-S}.
Incompleteness is due to three factors: a few percent decrease for
moderately bright sources due to occasional source overlap; a large
decrease for faint sources near the detection threshold; and a few
percent increase for very faint sources due to source overlap or
sources that lie in regions of particularly low background.  The
completeness curves shift as one moves off-axis because of the
radially degrading sensitivity limit of the CDFs.
Likewise, deviations in the recovered flux correction are due to three
factors: a few percent increase at all fluxes due to an additional
aperture correction not originally accounted for in A03; a steady
increase for faint sources due to photometry errors as detailed in
Figure~\ref{fig:flux_bias_explain}; and a steady decrease for faint
sources due to Eddington bias.  Eddington bias only appears to win out
over photometry effects for soft-band sources near the centers of the
fields.

\section{Number Counts}\label{sec:number_counts}

\begin{figure*}
\vspace{-0.1in}
\centerline{
\includegraphics[width=9.0cm]{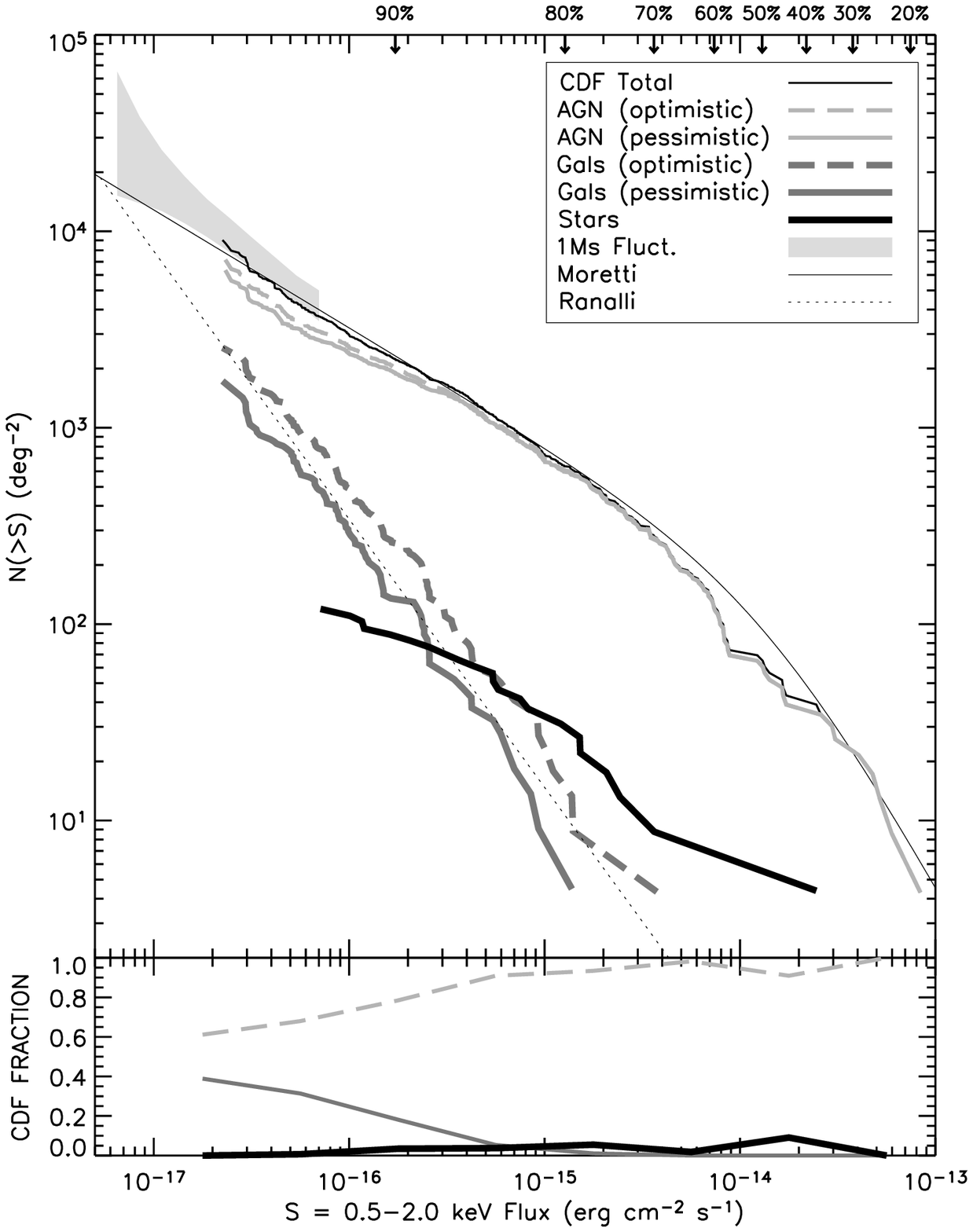}\hfill
\includegraphics[width=9.0cm]{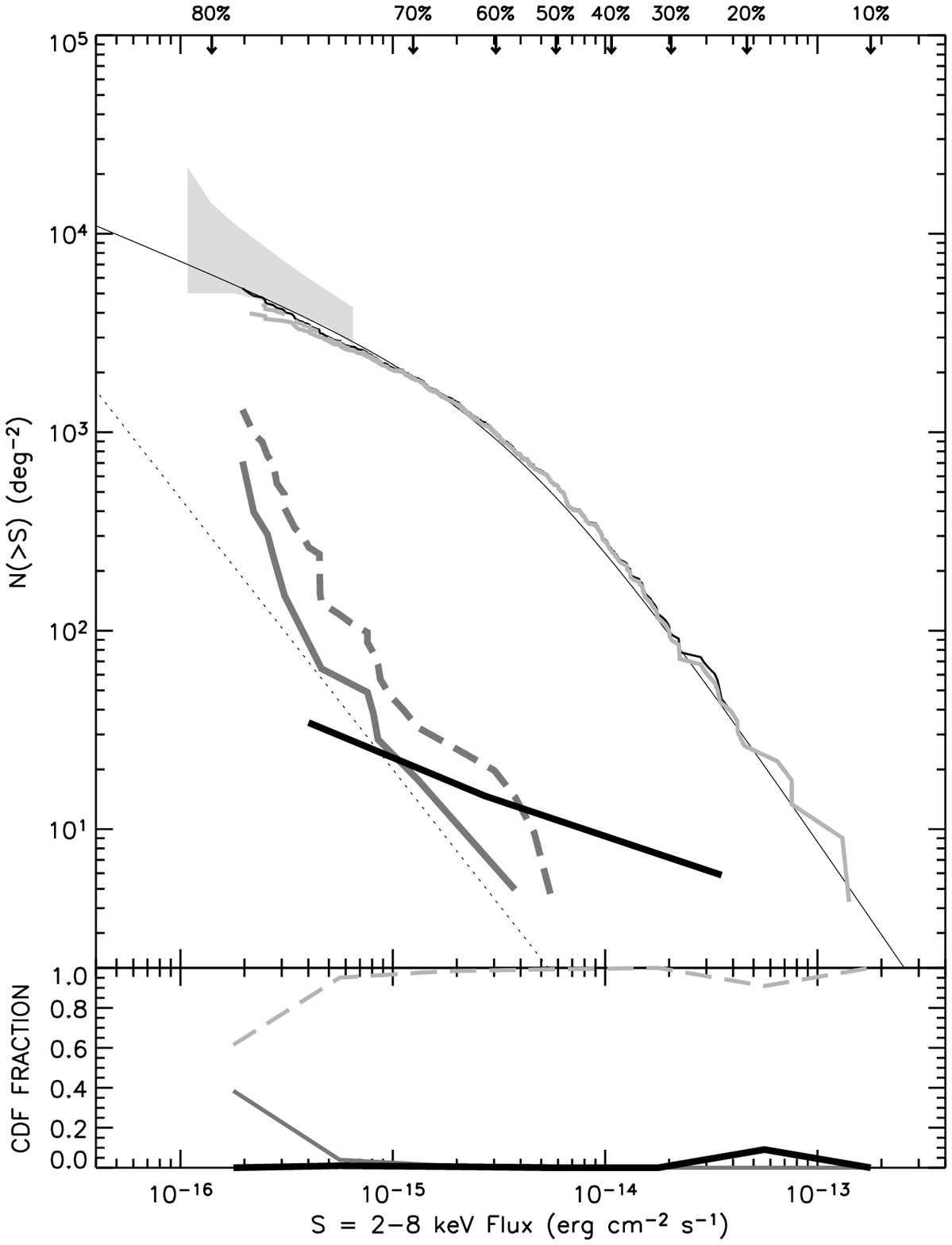}
}
\vspace{-0.1cm} \figcaption[Fig.4]{
  The soft ({\it left}) and hard ({\it right}) number counts in the
  CDFs are shown in the upper panels for all sources
  (solid black curves) and three different X-ray subsets: AGN (solid
  and dashed grey curves), galaxies (solid and dashed dark grey
  curves), and stars (thick solid black curves).  Also shown are the
  best-fit X-ray number count models of M03 (thin solid black curves),
  the 1~Ms \hbox{CDF-N} fluctuation analysis results from \citet[][grey
  ``fishtails'']{Miyaji2002}, and the predicted X-ray counts of
  star-forming galaxies from \citet{Ranalli2003}, calculated assuming
  the X-ray/radio correlation \citep[e.g.,][]{Bauer2002b, Ranalli2003}
  and the total (predominantly star-forming galaxies) radio number
  counts of \citet[][thin dotted grey lines]{Richards2000}. The
  percentage of the XRB resolved at a given X-ray flux
  \citetext{calculated from the number count models of M03, although
    renormalized in the hard band to the total hard XRB flux density
    of \citealp{DeLuca2004}} is shown at the top of the panels. Lower
  panels show the differential fraction of the total CDF soft- and
  hard-band samples each source type comprises.
\label{fig:num_cnts_agn_gal}}
\vspace{-0.3cm}
\end{figure*}

\begin{figure}
\vspace{-0.1in}
\centerline{
\includegraphics[width=9.0cm]{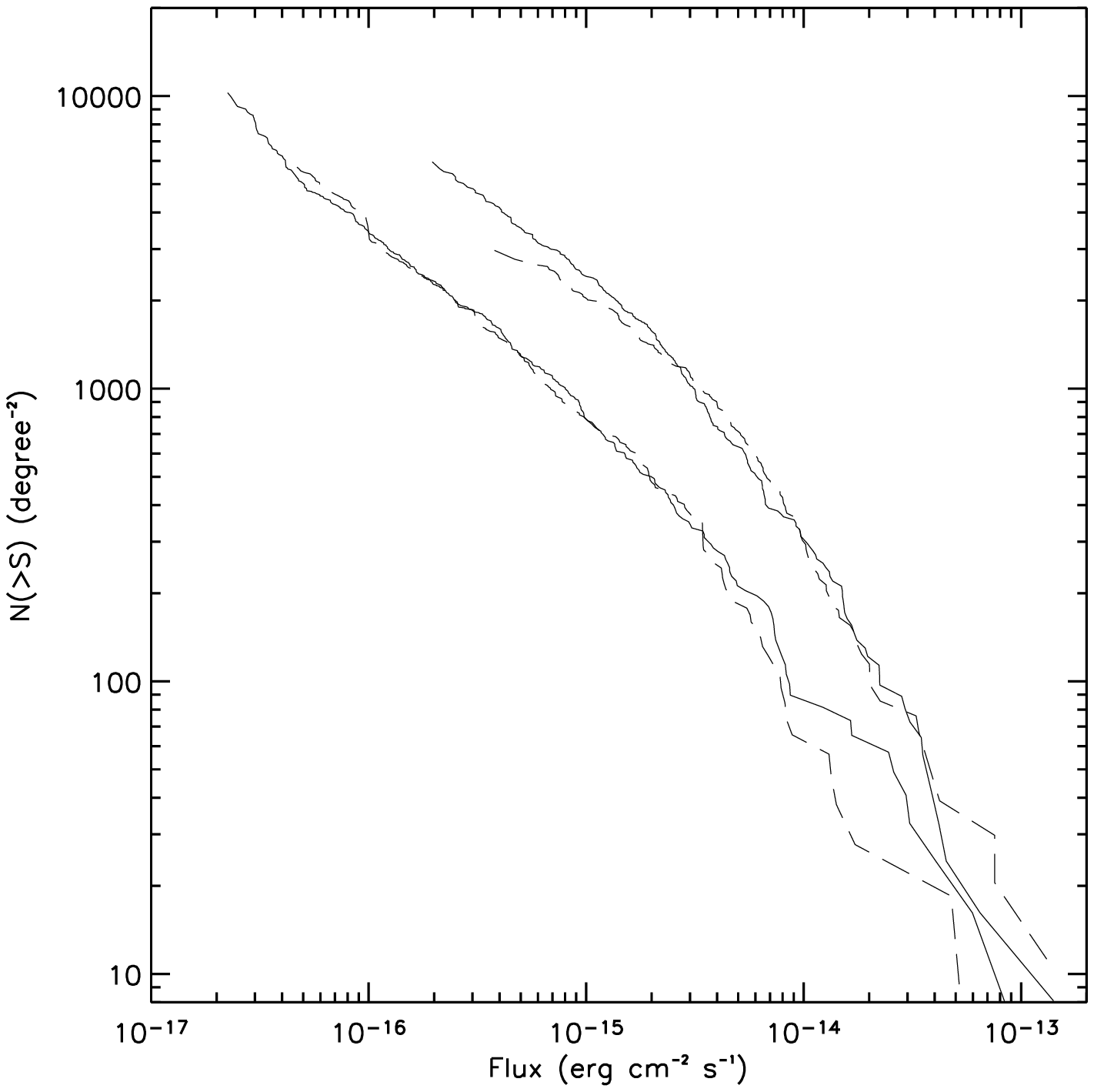}
}
\vspace{-0.1cm} \figcaption[Fig.5]{
  Comparison of \hbox{CDF-N} (solid curves) and \hbox{CDF-S} number
  counts (dashed curves) for the soft (lower set) and hard (upper set)
  bands. The two fields are consistent at the 1$\sigma$ confidence
  level over the entire flux range sampled in the soft band and above
  $F_{\rm 2-8~keV}\approx1\times10^{-15}$~erg~cm$^{-2}$~s$^{-1}$ in
  the hard band.
\label{fig:num_cnts_compare}}
\vspace{-0.3cm}
\end{figure}

The cumulative flux distribution ($\log N$-$\log S$) at each flux $S$,
for all sources brighter than $S$ weighted by the corresponding sky
coverage, is
\begin{equation}
N(>S)=\sum_{i=S_{i}>S} (CF_{i}~\Omega_{i})^{-1}, 
\end{equation}
where the sky coverage $\Omega_{i}$ is the maximum solid angle over
which each source could have been detected in both CDFs based on the
2$\sigma$ limiting flux maps, and $CF_{i}$ is the completeness
correction interpolated from the position- and count-dependent
simulated completeness correction (see $\S$\ref{sec:simulations}).
Additionally, each flux $S$ has been corrected for flux bias assuming
\begin{equation}
S_{i}=FR_{i}~S^{\rm o}_{i}, 
\end{equation}
where $FR_{i}$ is the position- and count-dependent simulated flux
recovery function (see $\S$\ref{sec:simulations}) and $S^{\rm o}_{i}$
is the original flux.  Figure~\ref{fig:num_cnts_agn_gal} shows the
total $\log N$-$\log S$ in the soft and hard bands for the combined
sample.  The combined $\log N$-$\log S$ curves are consistent with the
distributions and models found by other authors (e.g., C02; M03)
within the uncertainties. The CDFs appear to underestimate the
soft-band number counts around $F_{\rm
  0.5-2.0~keV}\approx10^{-14}$~erg~cm$^{-2}$~s$^{-1}$ compared to the
model curve of M03, although this is only a $\approx2\sigma$
deviation.  Figure~\ref{fig:num_cnts_compare} compares the number
counts in the \hbox{CDF-N} and \hbox{CDF-S}. We have worked out error
bars on the cumulative distributions following \citet{Gehrels1986} and
calculated the deviation in quadrature at each data point in units of
sigma.  We find that the number counts from the two fields are
consistent with each other at better than the $1\sigma$ confidence
level over the entire soft band and above $F_{\rm
  2-8~keV}\approx1\times10^{-15}$~erg~cm$^{-2}$~s$^{-1}$ in the hard
band. Below this hard-band flux, statistical deviations gradually
increase to $3.9\sigma$ at the faintest flux levels, a finding similar
to that presented in C02. This demonstrates that the field-to-field
variations previously reported for the CDFs are entirely consistent
with the lack of field-to-field variations found from ChaMP
\citep{Kim2004}, as the latter study only examined the X-ray $\log
N$-$\log S$ for relatively bright sources where we find the CDFs are
compatible.

The total number counts in both bands have previously been best fitted
with broken power laws with break fluxes between
$10^{-15}$--$10^{-14}$~erg~cm$^{-2}$~s$^{-1}$ (e.g., C02; M03). Given
this break range and the limited statistics of the CDFs above
$\sim10^{-14}$ erg~cm$^{-2}$~s$^{-1}$, we estimate the slope only for
the faint end of the number counts below $2\times10^{-15}$
erg~cm$^{-2}$~s$^{-1}$. We have assumed a single power-law model of
the form $N(>S)=N_{\rm 16} (S_{\rm i}/10^{-16})^{-\alpha}$, where the
slopes and normalizations were determined with a maximum-likelihood
algorithm \citep[e.g.,][]{Murdoch1973} using the sky-coverage
corrected (i.e., $CF_{i} \Omega_{i}$), differential flux
distribution. The best values of $\alpha$ were estimated by minimizing
\begin{eqnarray}
\nonumber L=M\ln\alpha - M\ln\sum_{i} (CF_{i} \Omega_{i}) (S_{i}^{-\alpha} - S_{i-1}^{-\alpha})\\
+ \sum_{i} \ln (CF_{i} \Omega_{i}) - (\alpha + 1)\sum_{i} \ln S_{i},
\end{eqnarray}
\noindent where $M$ is the number of sources used. The best values of
$N_{\rm 16}$ for a particular $\alpha$ were estimated by
\begin{eqnarray}
N_{\rm 16}= \frac{M}{\sum_{i} (CF_{i} \Omega_{i}) (S_{i}^{-\alpha} - S_{i-1}^{-\alpha})}.
\end{eqnarray}
The best-fit slopes and accompanying normalizations are provided in
Table~\ref{tab:slopes}. We typically do not have sufficient statistics
to ascertain whether a single power law appropriately represents the
underlying flux distribution; however, visual inspection of
Figure~\ref{fig:num_cnts_agn_gal} and the analyses below suggest that
such a simple model is unlikely to represent adequately sources near
or below our detection threshold due to the varied nature of the
contributing source populations.

\begin{deluxetable*}{ll|rr|rr|rr}
\tabletypesize{\footnotesize}
\tablecaption{Number Count Slopes and Normalizations\label{tab:slopes}} 
\tablehead{
\multicolumn{1}{c}{(1)} & 
\multicolumn{1}{c}{(2)} & 
\multicolumn{2}{c}{(3)} & 
\multicolumn{2}{c}{(4)} & 
\multicolumn{2}{c}{(5)}
\\
\multicolumn{1}{c}{Class} & 
\multicolumn{1}{c}{Subclass} & 
\multicolumn{2}{c}{\# } & 
\multicolumn{2}{c}{Slope ($\alpha$)} &
\multicolumn{2}{c}{$N_{\rm 16}$}
\\
\colhead{} & 
\colhead{} & 
\multicolumn{1}{c}{S} & 
\multicolumn{1}{c}{H} &
\multicolumn{1}{c}{S} & 
\multicolumn{1}{c}{H} &
\multicolumn{1}{c}{S} & 
\multicolumn{1}{c}{H}
}
\startdata
All           & Total         &    724 & 520 &  $0.55^{+0.03}_{-0.03}$ & $0.56^{+0.14}_{-0.14}$ & $3039^{ +88}_{-108}$ & $7403^{+125}_{-599}$ \\
\hline
AGN           & Total (opt.)  &    599 & 484 &  $0.47^{+0.03}_{-0.04}$ & $0.48^{+0.15}_{-0.15}$ & $2625^{+125}_{ -96}$ & $6901^{+306}_{-382}$ \\
              & Total (pess.) &    545 & 466 &  $0.41^{+0.03}_{-0.04}$ & $0.35^{+0.16}_{-0.16}$ & $2365^{+148}_{-104}$ & $6749^{+163}_{-201}$ \\
              & Optical Type~1&     73 &  65 &  $0.22^{+0.29}_{-0.22}$ & ---                    & $ 427^{+150}_{-103}$ & ---                    \\
      & Optical ``Not Type~1''&    526 & 419 &  $0.47^{+0.04}_{-0.03}$ & $0.37^{+0.10}_{-0.10}$ & $2266^{ +88}_{-106}$ & $4207^{+495}_{-97}$ \\
              & X-ray Unabs.  &    230 & 142 &  $0.20^{+0.05}_{-0.07}$ & $0.95^{+0.20}_{-0.21}$ & $ 998^{+359}_{-152}$ & $1868^{+444}_{-314}$ \\
              & X-ray Abs.    &    368 & 342 &  $0.62^{+0.04}_{-0.04}$ & $0.44^{+0.13}_{-0.11}$ & $1578^{ +59}_{ +60}$ & $4337^{+174}_{-117}$ \\
Galaxies      & Total (opt.)  &    157 &  30 &  $1.26^{+0.07}_{-0.06}$ & $2.02^{+0.33}_{-0.31}$ & $ 438^{ +17}_{ -19}$ & $4081^{+1358}_{-2033}$ \\
              & Total (pess.) &    103 &  12 &  $1.33^{+0.10}_{-0.08}$ & $2.51^{+0.53}_{-0.48}$ & $ 263^{ +16}_{ -18}$ & $3673^{+1671}_{-3038}$ \\
        & Starbursts (pess.)  &     55 &   5 &  $1.70^{+0.15}_{-0.12}$ & ---                    & $ 108^{ +10}_{ -12}$ & ---                    \\
              & Quiescent Gal.&     35 &   5 &  $1.30^{+0.17}_{-0.19}$ & ---                    & $ 115^{ +13}_{ -11}$ & ---                    \\
        & Ellipticals (pess.) &     13 &   2 &  $1.13^{+0.31}_{-0.31}$ & ---                    & $  56^{ +20}_{ -13}$ & ---                    \\
Stars         & Total         &     22 &   3 &  $0.65^{+0.12}_{-0.10}$ & ---                    & $ 125^{ +11}_{ -13}$ & ---                   \\
\hline
$\log(L_{\rm X})$ & $>44.5$   &     23 &  24 &  ---                    & ---                    & ---                  & ---                   \\
                 & 43.5--44.5 &    165 & 169 &  $0.11^{+0.02}_{-0.03}$ & $0.44^{+0.26}_{-0.44}$ & $ 780^{+102}_{ -91}$ & $1553^{+177}_{-200}$ \\
                 & 42.5--43.5 &    276 & 229 &  $0.43^{+0.04}_{-0.05}$ & $0.31^{+0.14}_{-0.13}$ & $1380^{+106}_{ -77}$ & $2533^{+728}_{-147}$ \\
                 & 41.5--42.5 &    115 &  57 &  $1.29^{+0.09}_{-0.08}$ & $1.46^{+0.20}_{-0.20}$ & $ 371^{ +19}_{ -23}$ & $4026^{+1144}_{-854}$ \\
                 & 40.5--41.5 &     17 &   4 &  $1.52^{+0.24}_{-0.23}$ & ---                    & $  48^{  +8}_{  -8}$ & ---                   \\
\hline
$\log(N_{\rm H})$ & 23--24    &    100 & 142 &  $0.96^{+0.08}_{-0.08}$ & $0.33^{+0.18}_{-0.33}$ & $ 381^{ +21}_{ -19}$  & $1619^{ +92}_{ -50}$ \\
                  & 22--23    &    250 & 180 &  $0.51^{+0.05}_{-0.05}$ & $0.60^{+0.16}_{-0.17}$ & $1097^{ +74}_{ -55}$  & $2974^{+341}_{-124}$ \\
                  & 21--22    &    145 & 103 &  $0.10^{+0.05}_{-0.10}$ & $0.98^{+0.22}_{-0.22}$ & $ 564^{+204}_{-163}$  & $2231^{+593}_{-411}$ \\
                  & <21       &    724 &  55 &  $0.54^{+0.09}_{-0.08}$ & $1.14^{+0.33}_{-0.35}$ & $ 483^{ +44}_{ -39}$  &  $985^{+500}_{-305}$ \\
\enddata
\tablecomments{ 
The two entries in columns 3--5 denote the soft (S) and hard (H) band
results. 
Column 1: Source classification (see $\S$\ref{sec:classification}).
Column 2: Source subsample (see $\S$\ref{sec:num_cnts_agn} and $\S$\ref{sec:num_cnts_gals}).
Column 3: Number of sources in class in the soft and hard samples,
respectively.
Columns 4 and 5: Number-count slopes ($\alpha$) and normalizations
($N_{\rm 16}$, sources~deg$^{-2}$ at $10^{-16}$~erg~cm$^{-2}$~s$^{-1}$) as
estimated using the maximum likelihood method
\citep[e.g.,][]{Murdoch1973} on the soft and hard samples,
respectively. Fitting was performed using only the sources with fluxes
below $2\times10^{-15}$~erg~cm$^{-2}$~s$^{-1}$ (i.e., below the known
break in the X-ray number counts) for all classes except that of
``Stars''.  For some cases, there were too few sources or too much
scatter below $2\times10^{-15}$~erg~cm$^{-2}$~s$^{-1}$ to provide
reliable number-count parameters. Such cases are denoted by ``---''.}
\end{deluxetable*}

Our best-fit slopes to the total faint-end number counts are
$0.55^{+0.03}_{-0.03}$ (soft) and $0.56^{+0.14}_{-0.14}$ (hard). These
values are slightly lower than previous estimates in both the soft
\citetext{e.g.,
0.70$\pm0.20$ from \citealp{Mushotzky2000}, 
0.67$\pm0.14$ from \citealp{Brandt2001b},
0.60$\pm0.10$ from \citealp{Rosati2002}, 
and 0.60$^{+0.02}_{-0.03}$ from M03} 
and hard \citetext{e.g., 
0.61$\pm0.10$ from \citealp{Rosati2002}, 
$0.63\pm0.05$ from C02, 
and 0.44$^{+0.12}_{-0.13}$ from M03} 
bands, although they are consistent within measurement errors. Our
lower values may be due to a number of the factors: (1) we are using
data which probe a factor of $\approx2$ deeper (i.e., the 2~Ms \hbox{CDF-N})
where the number counts may become flatter, (2) we estimate the
faint-end slope only using sources below
$2\times10^{-15}$~erg~cm$^{-2}$~s$^{-1}$ (i.e., a factor of several
lower than the break flux adopted by other studies), and (3) we use a
somewhat different technique to calculate the incompleteness and
Eddington bias corrections. The best-fitted number-count models from
M03 are also shown in Figure~\ref{fig:num_cnts_agn_gal} and display
excellent agreement apart from a slight soft-band deficit in the CDF
number counts around $F_{\rm
  0.5-2.0~keV}\sim10^{-14}$~erg~cm$^{-2}$~s$^{-1}$ (presumably due to
the effects of large-scale structure). At the CDF flux limits, we find
total source densities of 9014$^{+347}_{-334}$ and
5303$^{+248}_{-237}$ sources~deg$^{-2}$ in the soft and hard bands,
respectively (see Table~\ref{tab:stats}).

\begin{deluxetable*}{ll|rr|rr|rr}
\tabletypesize{\footnotesize}
\tablecaption{Number Count Statistics\label{tab:stats}} 
\tablehead{
\multicolumn{1}{c}{(1)} & 
\multicolumn{1}{c}{(2)} & 
\multicolumn{2}{c}{(3)} & 
\multicolumn{2}{c}{(4)} & 
\multicolumn{2}{c}{(5)}  
\\
\multicolumn{1}{c}{Class} & 
\multicolumn{1}{c}{Subclass} & 
\multicolumn{2}{c}{Total \# deg$^{-2}$} &
\multicolumn{2}{c}{CDF  XRB \%} &
\multicolumn{2}{c}{TOTAL XRB \%} 
\\
\colhead{} & 
\colhead{} & 
\multicolumn{1}{c}{S} & 
\multicolumn{1}{c}{H} &
\multicolumn{1}{c}{S} & 
\multicolumn{1}{c}{H} &
\multicolumn{1}{c}{S} & 
\multicolumn{1}{c}{H} 
}
\startdata
All           & Total         & 9014$^{+347}_{-334}$ & 5303$^{+248}_{-237}$ & 70.3$^{+4.4}_{-4.1}$ & 75.4$^{+5.3}_{-4.9}$ & 89.5$^{+5.9}_{-5.7}$ & 86.9$^{+6.6}_{-6.3}$ \\
\hline
AGN           & Total (opt.)  & 7166$^{+304}_{-292}$ & 4558$^{+216}_{-207}$ & 66.0$^{+4.3}_{-4.0}$ & 73.0$^{+5.1}_{-4.8}$ & 74.6$^{+5.9}_{-5.6}$ & 83.2$^{+6.5}_{-6.2}$ \\
              & Total (pess.) & 6342$^{+283}_{-271}$ & 3970$^{+192}_{-183}$ & 64.6$^{+4.3}_{-4.0}$ & 71.5$^{+5.1}_{-4.7}$ & 73.2$^{+5.9}_{-5.6}$ & 81.8$^{+6.5}_{-6.2}$ \\
              & Optical Type~1&  325$^{ +42}_{ -38}$ &  350$^{ +49}_{ -43}$ & 33.8$^{+4.7}_{-4.2}$ & 20.6$^{+3.1}_{-2.7}$ & 41.5$^{+6.2}_{-5.8}$ & 28.5$^{+5.0}_{-4.8}$ \\
         & Optical ``Not Type~1''& 6840$^{+311}_{-298}$ & 4207$^{+215}_{-205}$ & 32.2$^{+2.1}_{-2.0}$ & 52.4$^{+3.8}_{-3.5}$ & 33.1$^{+4.5}_{-4.5}$ & 54.7$^{+5.5}_{-5.3}$ \\
              & X-ray Unabs.  & 1774$^{+124}_{-116}$ & 1430$^{+130}_{-119}$ & 47.8$^{+4.1}_{-3.8}$ & 29.8$^{+3.1}_{-2.9}$ & 55.5$^{+5.7}_{-5.5}$ & 37.7$^{+5.1}_{-4.9}$ \\
              & X-ray Abs.    & 5392$^{+295}_{-280}$ & 3127$^{+178}_{-169}$ & 18.1$^{+1.3}_{-1.2}$ & 43.2$^{+3.3}_{-3.1}$ & 19.1$^{+4.2}_{-4.2}$ & 45.5$^{+5.2}_{-5.1}$ \\
Galaxies      & Total (opt.)  & 2552$^{+220}_{-203}$ & 1298$^{+283}_{-235}$ &  3.2$^{+0.3}_{-0.3}$ &  2.6$^{+0.6}_{-0.5}$ &  4.1$^{+4.0}_{-4.0}$ &  2.8~($<6.8$)        \\
              & Total (pess.) & 1727$^{+187}_{-169}$ &  711$^{+270}_{-202}$ &  1.8$^{+0.2}_{-0.2}$ &  1.2$^{+0.4}_{-0.3}$ &  2.7~($<6.7$)        &  1.4~($<5.6$)        \\
        & Starbursts (pess.)  &  984$^{+151}_{-132}$ &  123$^{ +84}_{ -53}$ &  0.9$^{+0.1}_{-0.1}$ &  0.4$^{+0.2}_{-0.2}$ &  ---                 &  ---                 \\
              & Quiescent Gal.&  343$^{ +68}_{ -57}$ &  481$^{+327}_{-207}$ &  0.6$^{+0.1}_{-0.1}$ &  0.7$^{+0.4}_{-0.3}$ &  ---                 &  ---                 \\
        & Ellipticals (pess.) &  399$^{+144}_{-109}$ &  105$^{+139}_{ -67}$ &  0.2$^{+0.1}_{-0.1}$ &  0.1$^{+0.2}_{-0.1}$ &  ---                 &  ---                 \\
Stars         & Total         &  119$^{ +31}_{ -25}$ &   34$^{ +33}_{ -18}$ &  2.5$^{+0.7}_{-0.5}$ &  1.3$^{+1.2}_{-0.7}$ & 10.2$^{+4.1}_{-4.0}$ &  1.4~($<5.6$)        \\
\hline
$\log(L_{\rm X})$ & $>44.5$   &  109$^{ +27}_{ -22}$ &  113$^{ +28}_{ -23}$ & 13.0$^{+3.4}_{-2.7}$ & 12.7$^{+3.2}_{-2.6}$ &  ---                 &  ---                 \\
                 & 43.5--44.5 &  868$^{ +73}_{ -67}$ & 1050$^{ +87}_{ -80}$ & 33.7$^{+3.3}_{-3.0}$ & 33.9$^{+3.3}_{-3.0}$ &  ---                 &  ---                 \\
                 & 42.5--43.5 & 2716$^{+173}_{-163}$ & 2157$^{+152}_{-142}$ & 15.9$^{+1.3}_{-1.2}$ & 22.1$^{+1.9}_{-1.8}$ &  ---                 &  ---                 \\
                 & 41.5--42.5 & 3055$^{+309}_{-281}$ & 1168$^{+176}_{-154}$ &  3.1$^{+0.3}_{-0.3}$ &  4.2$^{+0.7}_{-0.6}$ &  ---                 &  ---                 \\
                 & 40.5--41.5 &  411$^{+126}_{ -98}$ &   58$^{ +46}_{ -28}$ &  2.6$^{+0.8}_{-0.6}$ &  1.7$^{+1.4}_{-0.8}$ &  ---                 &  ---                 \\
\hline
$\log(N_{\rm H})$ & 23--24    & 1754$^{+193}_{-175}$ & 1174$^{+107}_{ -98}$ &  3.0$^{+0.4}_{-0.3}$ & 17.2$^{+1.8}_{-1.7}$ &  ---                 &  ---                 \\
                  & 22--23    & 3452$^{+232}_{-218}$ & 1668$^{+133}_{-124}$ & 14.3$^{+1.2}_{-1.1}$ & 23.9$^{+2.3}_{-2.1}$ &  ---                 &  ---                 \\
                  & 21--22    &  862$^{ +77}_{ -71}$ & 1199$^{+130}_{-117}$ & 18.4$^{+1.9}_{-1.7}$ & 14.8$^{+1.8}_{-1.6}$ &  ---                 &  ---                 \\
                  & $<21$     & 1098$^{+118}_{-107}$ &  435$^{ +67}_{ -58}$ & 30.2$^{+3.6}_{-3.2}$ & 16.7$^{+2.7}_{-2.4}$ &  ---                 &  ---                 \\
\enddata
\tablecomments{ 
The two entries in columns 3--6 denote the soft (S) and hard (H) band results.  
Column 1: Source classification (see $\S$\ref{sec:classification}).
Column 2: Source subsample (see $\S$\ref{sec:num_cnts_agn} and $\S$\ref{sec:num_cnts_gals}).
Column 3: Source number density in the soft and hard samples,
respectively.
Column 4: Percentage of the XRB flux density which each source class
contributes (considering detected CDF sources only).
Column 5: Percentage of the XRB flux density which each source class
contributes (considering CDF sources and sources brighter than those
in the CDFs, which we estimate to comprise a total of 19.2\% in the
soft band and 11.5\% in the hard band; see
$\S$\ref{sec:number_counts}).  For some classes, the percentage of the
XRB for brighter sources is not well known and has been omitted
(denoted by ``---'').
}
\end{deluxetable*}

\begin{figure*}
\vspace{-0.1in} \centerline{
\includegraphics[width=9.0cm]{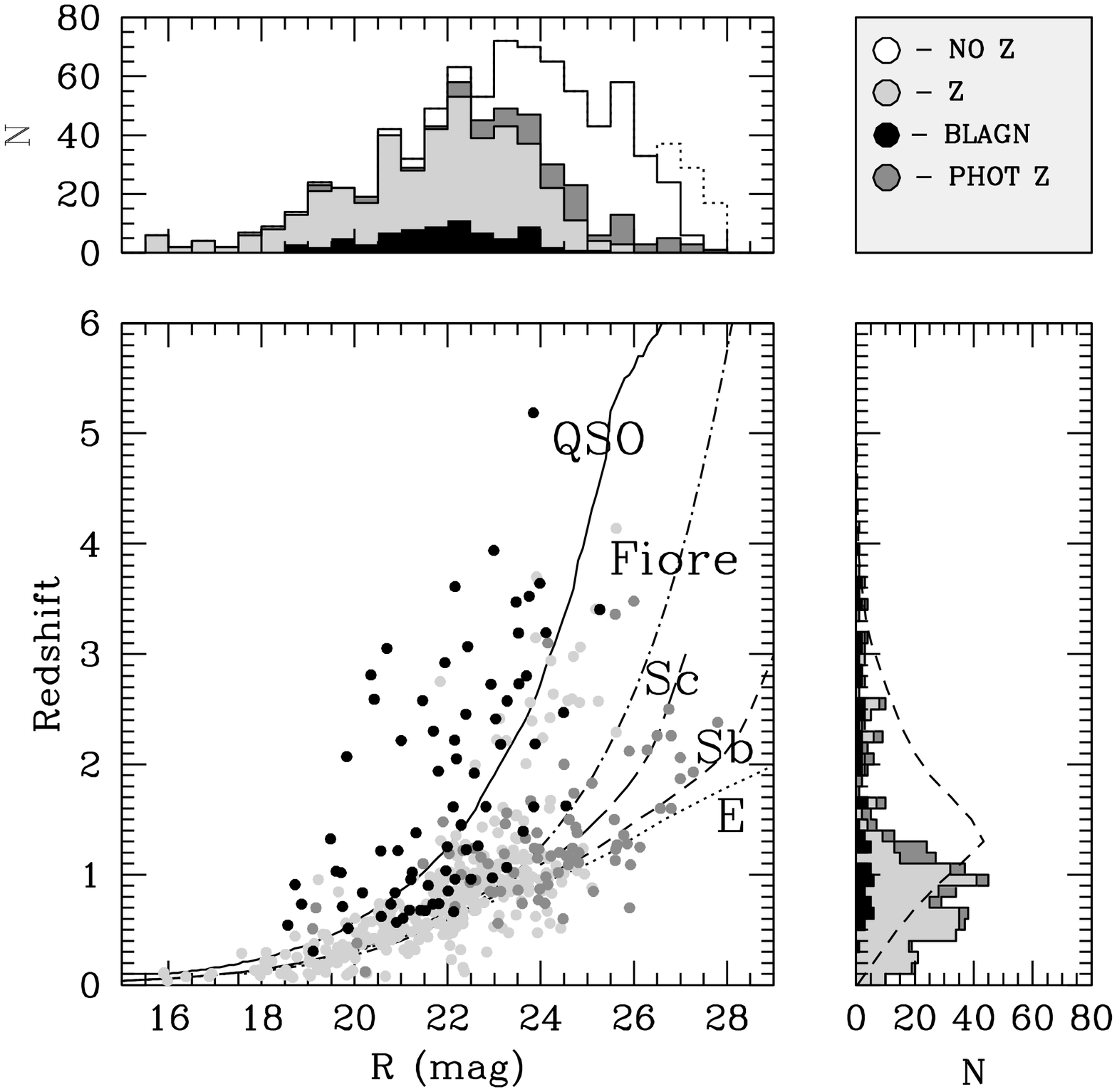}\hfill
\includegraphics[width=9.0cm]{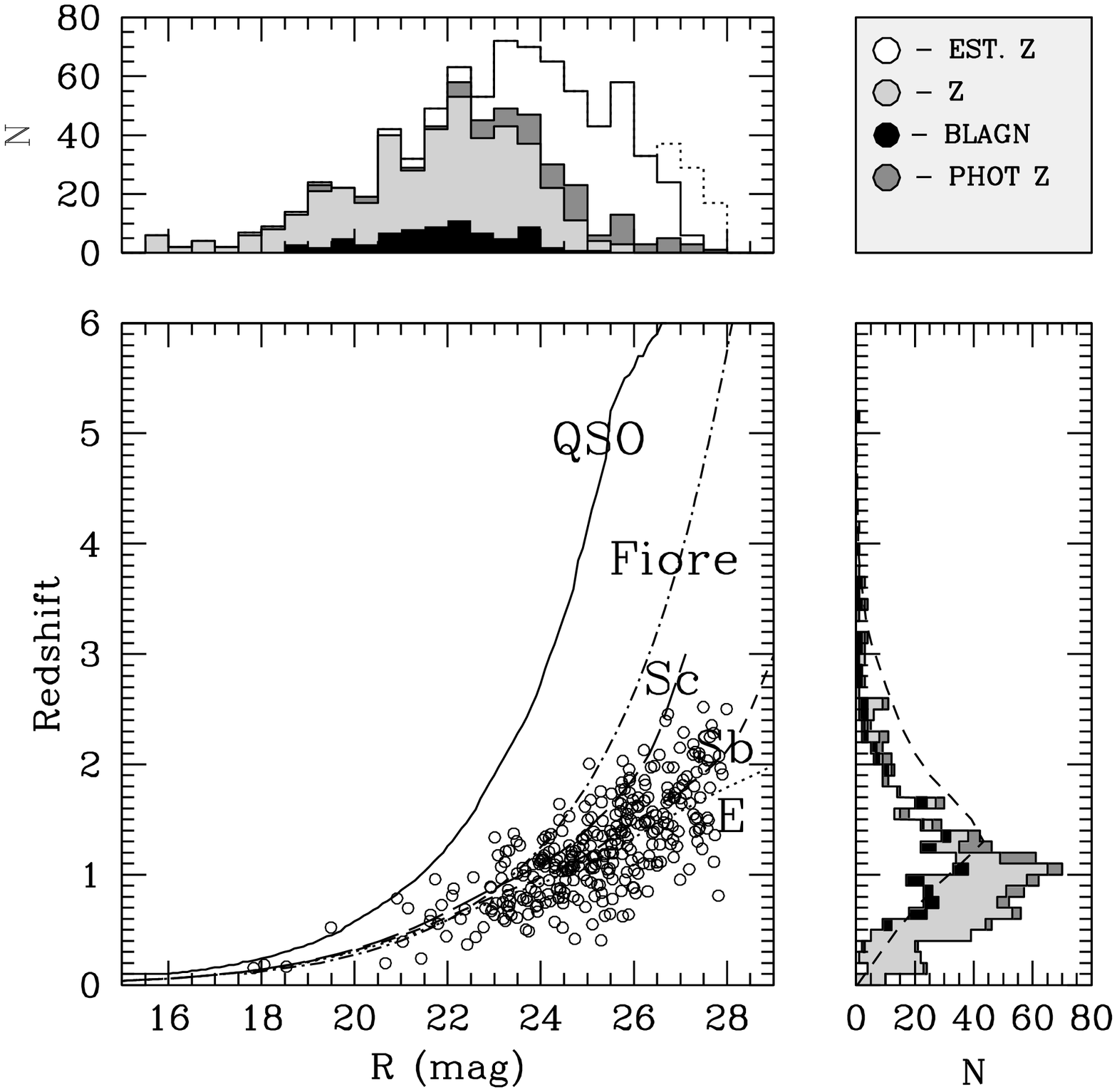}
} \vspace{-0.1cm} 
\figcaption[Fig.6]{ 
  The $R$-band magnitude versus redshift distribution for all 829 CDF
  sources, divided into sources with known redshifts ({\it left}) and
  unidentified CDF sources ({\it right}) adopting an
  $R$-band-to-redshift estimate based on an Sb galaxy spectral energy
  distribution \citep[see
  $\S$\ref{sec:classification};][]{Coleman1980}. The sources with
  known redshifts have additionally been subdivided into
  spectroscopically (Z) and photometrically (PHOT Z) identified
  objects, as well as those with broad emission lines (BLAGN).
  Templates for an $M_{\rm B}=-23$ QSO and unevolved $M^{*}$ Sc, Sb,
  and E galaxies are shown overlaid, adopting $K$-corrections based on
  a custom composite QSO spectrum \citep[consistent with that
  of][where they overlap]{VandenBerk2001} and the galaxy models of
  \citet{Poggianti1997}, respectively. For comparison, we also plot
  the correlation between intrinsic $L_{X}$ and $f_{\rm X}/f_{\rm O}$
  discussed by \citet[][]{Fiore2003}, which can be reduced to a
  correlation between optical magnitude and redshift provided we apply
  a $\approx$15\% correction factor to account for the typical
  intrinsic X-ray absorption these authors find in $L_{X}$ \citep[for
  details see][]{Fiore2003}. Additionally, we show corresponding
  $R$-band and redshift histograms to illustrate the overall
  distributions. The dotted histogram in the $R$-band magnitude
  distribution represent the 52 sources with no known redshift and no
  $R$-band counterpart (see $\S$\ref{sec:num_cnts_type} for details).
  The dashed curve in the redshift histogram indicates the expected
  distribution of AGN measured from {\it ROSAT} observations
  \citep{Gilli2003a}, normalized to the total number of AGN in the
  CDFs (assumed to be $\sim$80\% of all CDF sources; see column~3 of
  Table~\ref{tab:stats}). The large discrepancy between {\it ROSAT}
  and CDF sources at low redshift is due to the fact that CDF AGN
  appear to peak in number density at somewhat lower redshifts
  \citep[e.g., ][]{Barger2003b, Ueda2003}. There is also a
  non-negligible and increasing fraction of star-forming galaxies in
  the CDF sample \citep[e.g., ][]{Hornschemeier2003a}.
\label{fig:R_vs_z}}
\vspace{-0.3cm}
\end{figure*} 

\subsection{Number Counts by Type}\label{sec:num_cnts_type}

A large fraction of the X-ray-bright CDF sources can be securely
identified as AGN based on their X-ray luminosities, X-ray spectral
properties, radio properties, optical spectroscopic classifications,
and X-ray-to-optical flux ratios \citep[e.g.][]{Alexander2002b,
  Bauer2002b}. However, the classifications of many faint X-ray
sources remain ambiguous because they lack firm redshifts, and their
X-ray properties are consistent with emission from either a
low-luminosity active nucleus, star formation, or a combination of
both. To understand these sources, we must make a few reasonable
assumptions about the nature of the CDF sources as outlined below.
Following these assumptions, we adopt two classification schemes,
one which conservatively estimates the number AGN and one which
conservatively estimates the number of star-forming X-ray sources.

Optical magnitudes for the CDF sources have been measured using Subaru
observations in the \hbox{CDF-N} \citep{Barger2003b} and Wide-Field Imager
(WFI) observations as part of an ESO deep public survey in the \hbox{CDF-S}
\citep[e.g.,][]{Arnouts2001}.  Fifty-two of the CDF sources have no
optical counterpart to the depths of the Subaru and WFI images
($R\ga27$). We have assigned $R$-band magnitudes to these blank X-ray
sources assuming they represent the tail of the currently observed
$R$-band counterparts. This assumption is validated by deeper {\it
  HST} imaging for a large subset of the CDF sources \citep[e.g.,
Bauer et al. 2004, in preparation;][]{Koekemoer2004}, although we
caution that a small fraction of sources could be much fainter than
our estimates. Redshifts for the X-ray sources were culled from
several recent spectroscopic \citep{Barger2003b, Steidel2003,
  LeFevre2004, Szokoly2004, Wirth2004} and photometric
\citep{Alexander2001, Barger2003b, Wolf2004} catalogs.  Of the 829 CDF
sources, 425 have spectroscopic redshifts and another 80 have firm
photometric redshifts. The remaining 324 sources that lack redshifts
are nearly all optically faint ($R\ga24$) and have proven difficult to
identify in large numbers even on 8--10m class telescopes. In order to
classify all sources, we first must estimate the redshifts for these
remaining sources (thereby allowing X-ray luminosity and rest-frame
absorption column determinations).  We note that detailed studies of
optically faint X-ray sources indicate that they often have hard X-ray
colors typical of obscured AGN and red optical colors typical of
early-type galaxies at $z\sim1$--3 \citep[e.g., ][]{Alexander2001,
  Koekemoer2004, Treister2004}.

In Figure~\ref{fig:R_vs_z}, we show the distribution of redshift
versus $R$-band magnitude for the CDF sources with known redshifts, as
well as templates for an $M_{\rm B}=-23$ QSO and unevolved $M^{*}$ Sc,
Sb, and E galaxies. $K$-corrected galaxy tracks were derived from the
study of \citet{Poggianti1997}. Since no evolutionary corrections have
been made, these tracks should be considered extreme in the sense that
a typical $z=1$ elliptical should have bluer colors than those shown
here. The $K$-corrected QSO track was calculated using a custom
composite QSO spectrum \citep[consistent with that of][where they
overlap]{VandenBerk2001} assuming the QSO continuum has $\alpha=0.5$
(where $F_{\nu}\propto\nu^{-\alpha}$), typical emission-line
strengths, and standard absorption due to the Lyman alpha forest.

While the broad-line AGN span a wide range in $R$ and are generally
consistent with the QSO track, the overwhelming majority of the other
CDF sources follow the galaxy tracks, indicating that their optical
light is likely to be dominated by their host galaxies \citetext{see
  \citealp{Grogin2003} for confirmation of this dominance of the
  optical host galaxies from {\it HST} imaging}. We note that the
correlation found by \citet{Fiore2003} between intrinsic $L_{X}$ and
$f_{\rm X}/f_{\rm O}$ can essentially be reduced to one between
optical magnitude and redshift (such as is seen in
Figure~\ref{fig:R_vs_z}) as long as we include a correction for X-ray
absorption, which these authors suggest is typically $\la$15\% for
$\approx$90\% of their sources and rarely excedes factors of a few. In
this new form, the \citet{Fiore2003} correlation tracks bright optical
sources as well as the galaxy templates, but appears to overestimate
redshifts for sources with $R>24$ by increasingly large factors.  Thus
caution should be exercised when extrapolating this correlation to
extremely faint optical magnitudes \citep[cf.][]{Padovani2004}.

The fainter sources that lack redshifts are likely to continue to
follow the galaxy tracks in Figure~\ref{fig:R_vs_z}, and thus in the
absence of reliable photometric redshifts we can use this good
correspondence as a crude redshift indicator to glean basic trends.
Differences between the galaxy templates do not appear large below
$R\sim27$.  However, the Sb template formally provides the best
empirical fit to the available data \citep[see also, e.g.,][]{Alexander2002a,
  Barger2002, Barger2003b} and was therefore adopted as our median
spectral energy distribution (SED) to convert $R$-band magnitude to
redshift. We assumed a scatter about this template equal to that
measured between $R$=23--24 for the known redshift distribution. We
caution that this technique is only valid in a statistical sense and
is only strictly true if (1) fainter X-ray sources remain host-galaxy
dominated at optical wavelengths and (2) their typical hosts are
well-represented by an Sb galaxy SED.  Reassuringly, the small number
of optically faint sources with redshift determinations follow this
template. 

We must also calculate rest-frame X-ray absorption column densities
for all CDF sources in order to determine the intrinsic power of the
AGN.  These were determined from direct X-ray spectral analysis (Bauer
et al. 2004, in preparation), which is more accurate than using simple
band ratios. An absorbed power-law model
\citep[$wabs+po$;][]{Morrison1983} was fitted to all CDF sources.  The
absorption and normalization were varied as free parameters. The
photon index of the power law was allowed to vary such that
$\Gamma\ga1.7$ for sources with more than 100 counts in the
0.5--8.0~keV band, while it was fixed at $\Gamma=1.7$ for sources
below 100 counts. The fitting was performed on the unbinned X-ray data
using the Cash statistic \citep{Cash1979} to maximize spectral
information for low-count sources.  The spectral parameters of the CDF
sources were then used to calculate unabsorbed, rest-frame
0.5--8.0~keV luminosities.

\subsubsection{Source Classification}\label{sec:classification}

Our first source classification scheme is based on intrinsic
0.5--8.0~keV luminosities, X-ray spectral properties, radio properties
(slopes, morphologies, and variability), and optical spectroscopic
classifications. Our specific AGN criteria are motivated by the
following findings. From X-ray observations in the local Universe,
purely star-forming galaxies do not appear to have X-ray luminosities
exceeding $L_{\rm 0.5-8.0~keV}\approx3\times10^{42}$~erg~s$^{-1}$, and
rarely, if ever, have intrinsic absorption column densities above
$N_{\rm H}\approx10^{22}$~cm$^{-2}$ \citep[e.g.,][]{Fabbiano1989,
  Colbert2004}. The above numbers are probably conservative,
considering the most X-ray-luminous local star-forming galaxy known,
NGC~3256, only has an X-ray luminosity of $L_{\rm
  0.5-8.0~keV}\la10^{42}$ \citep{Moran1999, Lira2002} and that X-ray
emission from star-forming galaxies is typically extended; obscuring
X-ray emission with an average absorption column density of $N_{\rm
  H}\ga10^{22}$~cm$^{-2}$ would require extraordinary amounts of
intervening gas.  Finally, nearby extragalactic X-ray sources with
extremely flat X-ray spectral slopes ($\Gamma<1$) are almost always
identified as highly obscured or Compton-thick AGN, where the primary
emission from the AGN is almost completely obscured and we only see
the flat scattered or reflected component
\citep[e.g.,][]{Maiolino1998, Bassani1999}. At radio wavelengths, the
vast majority of sources with powerful radio jets/lobes, flat radio
spectral slopes, or strong radio variability are AGN
\citep[e.g.,][]{Condon1984, Condon1986}, while extragalactic sources
with broad (EW$>$1000~km~s$^{-1}$) or high-excitation optical emission
lines are almost universally classified as AGN
\citep[e.g.,][]{Osterbrock1989}.

Based on the above constraints, the CDF sources were divided into AGN,
galaxies, and stars. Twenty-two sources were classified as Galactic
stars based on their spectroscopic identifications. Six hundred
thirty-two sources were classified as AGN based on at least one of the
following properties: $N_{\rm H}\ge10^{22}$~cm$^{-2}$, hardness ratio
$> 0.8$ (equivalent to effective $\Gamma<1.0$), $L_{\rm 0.5-8.0~keV} >
3\times10^{42}$~erg~s$^{-1}$, or broad/high-ionization AGN emission
lines. We recognize that there will be AGN-dominated sources not
selected by these criteria, such as AGN with $\Gamma>1.0$ and
rest-frame luminosities $L_{\rm 0.5-8.0~keV}<3\times
10^{42}$~erg~s$^{-1}$; however, such AGN are difficult to classify
even locally. Seventy-six sources were classified as star-forming
galaxies based on either having off-nuclear X-ray emission or $N_{\rm
  H} < 10^{22}$~cm$^{-2}$, hardness ratio $< 0.8$, and $L_{\rm
  0.5-8.0~keV} < 3\times10^{42}$~erg~s$^{-1}$. The classifications of
a remaining 99 CDF sources were considered ambiguous due to weak X-ray
spectral constraints; we have tentatively classified these sources as
star-forming galaxies, although there is likely to be some degree of
obscured and low-luminosity AGN contamination. Note that the numbers
provided above are based on the classification of all 829 CDF sources,
and should not be confused with the numbers given in
Table~\ref{tab:slopes} which indicate the number of each source type
used in the soft-band and hard-band number-count estimates,
respectively.

\begin{figure}
\vspace{-0.1in}
\centerline{
\includegraphics[width=9.0cm]{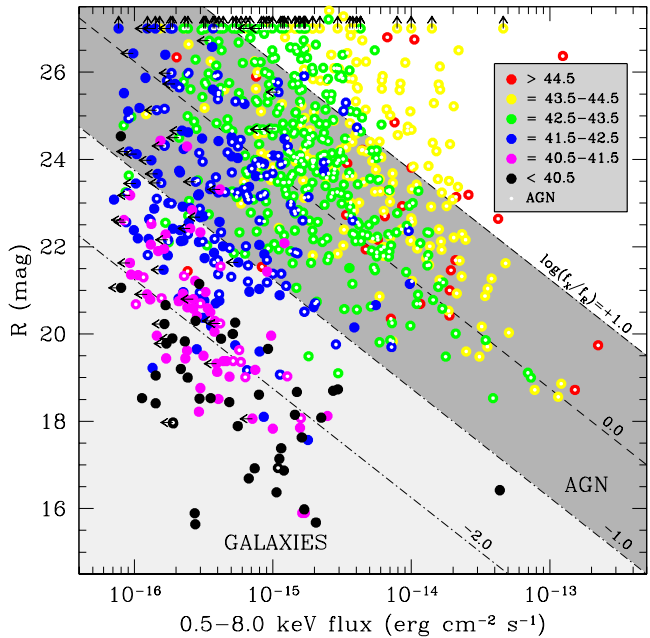}
}
\vspace{-0.1cm} \figcaption[Fig.7]{ 
  $R$-band magnitude versus 0.5--8.0~keV flux for all 829 CDF sources.
  Sources are divided into several estimated, unabsorbed 0.5--8.0~keV
  luminosity bins (color-coded following the order of the rainbow,
  with red denoting the highest luminosity and absorption bins).
  Sources that are considered AGN following the criteria in
  $\S$\ref{sec:classification} are denoted by small white dots.
\label{fig:R_vs_x}}
\vspace{-0.3cm}
\end{figure} 

Figure~\ref{fig:R_vs_x} compares $R$-band magnitudes and 0.5--8.0~keV
fluxes for all 829 CDF sources, separated into several unabsorbed,
rest-frame 0.5--8.0~keV luminosity classes. Those sources classified
as AGN under our first scheme are highlighted. From inspection, it is
apparent that the X-ray-to-optical flux ratio crudely tracks X-ray
luminosity (and hence AGN activity), explaining why $f_{\rm
  0.5-8.0~keV}/f_{\rm R}=0.1$ is a useful AGN/galaxy discriminator
\citep[e.g.,][]{Maccacaro1988, Stocke1991, Schmidt1998, Akiyama2000,
  Hornschemeier2000}. At very faint X-ray flux levels, however, the
relation will begin to break down for certain source types due to the
very different nature of their X-ray and optical K-corrections (most
notable in this regard are Type~2 Seyfert galaxies, whose
X-ray-to-optical flux ratios can vary by more than an order of
magnitude between $z=0$--2; Moran et al., in preparation).  There is
also likely to be some degree of obscured and low-luminosity AGN
contamination at $f_{\rm 0.5-8.0~keV}/f_{\rm R}<0.1$, as well as
considerable potential galaxy contamination for $f_{\rm
  0.5-8.0~keV}/f_{\rm R}>0.1$ and $F_{\rm
  0.5-8.0~keV}\la4\times10^{-16}$~erg~cm$^{-2}$~s$^{-1}$. We attribute
the fact that star-forming galaxies begin to have $f_{\rm
  0.5-8.0~keV}/f_{\rm R}>0.1$ at faint X-ray fluxes to (1) a stronger
K-correction in the optical band than in the X-ray band that tends to
shift potential star-forming galaxies into the AGN region and (2)
increasingly large uncertainties in the source band ratios (the
primary AGN discriminator for low-luminosity X-ray sources) at faint
fluxes.

We can estimate the prevalence of obscured AGN among these ambiguous
faint X-ray sources by stacking them in suitable subsets and examining
their average X-ray band ratios. Given the good correspondence between
X-ray-to-optical flux ratio and AGN activity, we stacked all of the
ambiguously classified X-ray sources with unconstrained band ratios
and $F_{\rm 0.5-8.0~keV}\le4\times10^{-16}$~erg~cm$^{-2}$~s$^{-1}$ in
decades of decreasing X-ray-to-optical flux ratio. This yielded
average effective $\Gamma$'s of 1.35, 1.45, 1.77, and 2.01 (associated
errors are $\pm$10--20\%) for $f_{\rm 0.5-8.0~keV}/f_{\rm R}=1$--10,
0.1--1, 0.01--0.1, and $<0.01$, respectively. The steep effective
photon indices for sources stacked in the ``galaxies'' region of
Figure~\ref{fig:R_vs_x} suggest that there are relatively few obscured
AGN among these sources and that they are likely star-forming in
nature (although we cannot exclude the presence of soft low-luminosity
AGN or AGN/starburst composites). The flatter effective photon indices
for sources in the traditional AGN region, on the other hand, suggest
that a significant fraction are in fact obscured AGN, with the rest
presumably a mix of unobscured AGN and star-forming galaxies. Thus,
our first classification scheme appears to provide a pessimistic
estimate of AGN and an optimistic estimate of star-forming galaxies
(since some of the ambiguous sources classified as star-forming
galaxies may in fact be powered by AGN).

Our second classification scheme is a slight variation of the first
and is similar to that presented in \citet{Alexander2002b} and
\citet{Bauer2002b}. In addition to the X-ray spectral properties,
intrinsic X-ray luminosities, radio morphologies, variability, and
optical spectroscopic classifications, we further classify ambiguous
sources based on their X-ray-to-optical flux ratios. Thus we
considered 698 sources to be AGN based on at least one of the
following properties: $f_{\rm 0.5-8.0~keV}/f_{\rm R} > 0.1$, $N_{\rm
  H}\ge10^{22}$~cm$^{-2}$, hardness ratio $> 0.8$ (equivalent to
effective $\Gamma<1.0$), $L_{\rm 0.5-8.0~keV} >
3\times10^{42}$~erg~s$^{-1}$, or broad/high-ionization AGN emission
lines. The remaining 109 sources were considered star-forming
galaxies, although again some low-luminosity AGN might be present. In
contrast to our first scheme, this division should provide a more
optimistic estimate of AGN and a more pessimistic estimate of
star-forming galaxies, as it attempts to address the issue of strong
AGN contamination for unclassified $f_{\rm 0.5-8.0~keV}/f_{\rm R}>0.1$
sources. 

Number counts for the two classification schemes are compared in
Figure~\ref{fig:num_cnts_agn_gal}, and our best-fit slopes to the
total faint-end number counts are provided in Table~\ref{tab:slopes}
for each category. Importantly, the slopes of the AGN and galaxy
number counts are not strongly affected by the adopted scheme. The
true number counts for each class are likely to lie somewhere in
between these two determinations. Given that the stacked effective
photon indices of the ambiguous sources in the AGN region are
relatively hard, and that there is likely to be some soft AGN
contamination in the ``galaxies'' region, we will adopt the second
scheme presented above (optimistic AGN, pessimistic galaxies).

We can compare our pessimistic galaxy sample to the Bayesian selected
sample of normal/starburst galaxies in the CDFs from
\citet{Norman2004}. Using only CDF sources with spectroscopic
redshifts $z<1.2$ and a selection based on X-ray luminosities,
observed X-ray hardness ratios, and X-ray-to-optical flux ratios,
\citet{Norman2004} identified a total of 210 normal/starburst
galaxies.  This number is substantially larger than the 109 galaxies
in our pessimistic galaxy sample or even the 175 galaxies in our {\it
  optimistic} sample and warrants investigation.  Part of this
discrepancy arises from the fact that \citet{Norman2004} use the
catalog of \citet{Giacconi2002} rather than that of A03; there are 18
\citet{Giacconi2002} sources in their sample which are not in the main
catalogs of A03. Of the remaining 192 \citet{Norman2004} galaxies, we
find 120 and 94 matches to our optimistic and pessimistic samples,
respectively. Of the 55 sources classified as galaxies here but not by
\citet{Norman2004}, $\sim30$\% were objects lacking spectroscopic
redshifts or with $z>1.2$ (15 cases) while it is unclear why
the rest failed their criteria (40 cases). Conversely, sources
classified as galaxies by \citet{Norman2004} but AGN here were objects
which had a combination of $f_{\rm 0.5-8.0~keV}/f_{\rm R} > 0.1$ (62
cases), $N_{\rm H}\ge10^{22}$~cm$^{-2}$ (44 cases), hardness ratio $>
0.8$ (23 cases), or $L_{\rm 0.5-8.0~keV} >
3\times10^{42}$~erg~s$^{-1}$ (20 cases). Based on our criteria, we
expect significant AGN contamination to be present in the
\citet{Norman2004} sample.

We note that the upward trend seen below $F_{\rm
  0.5-2.0~keV}\sim5\times10^{-17}$~erg~cm$^{-2}$~s$^{-1}$ in the AGN
number counts in Figure~\ref{fig:num_cnts_agn_gal} and elsewhere is
likely caused by obscured AGN which have composite X-ray spectra, such
that a significant fraction of the soft-band flux is likely due to
either star formation associated with the host galaxy or complex AGN
spectra (e.g., partial covering, scattered radiation, or reflection).

The AGN number-count slopes are flat in both bands, indicating that we
are now observing AGN beyond the differential peak in the AGN
number-count distribution.  While the X-ray source densities of AGNs,
in a differential sense (see lower panel of
Figure~\ref{fig:num_cnts_agn_gal}), appear to be in a state of decline
at X-ray fluxes below $\sim10^{-15}$~erg~cm$^{-2}$~s$^{-1}$, the
star-forming galaxy population is rising strongly \citep[nearly
Euclidean and consistent with the previously measured normal galaxy
X-ray number counts of][]{Hornschemeier2003a}. The star-forming galaxy
number counts also agree well with the predictions made by
\citet{Ranalli2003} using the X-ray/radio correlation
\citep[e.g.,][]{Bauer2002b, Ranalli2003} and the radio number counts
of \citet{Richards2000}, indicating that there is a good
correspondence between the X-ray and radio emission in these sources.

For completeness we also present the X-ray number counts for the 22
spectroscopically identified, Galactic stars discovered in the CDFs.
These high-latitude stars are typically old G, K, and M types which
are thought to emit X-ray emission via magnetic flaring; we refer the
reader to \citet{Feigelson2004}, who have performed analyses on a
subset of the X-ray-detected stars presented here. The number-count
slope of the Galactic stars is flat in the soft band and
indeterminate in the hard band.

\begin{figure*}
\vspace{-0.1in}
\centerline{
\includegraphics[width=9.0cm]{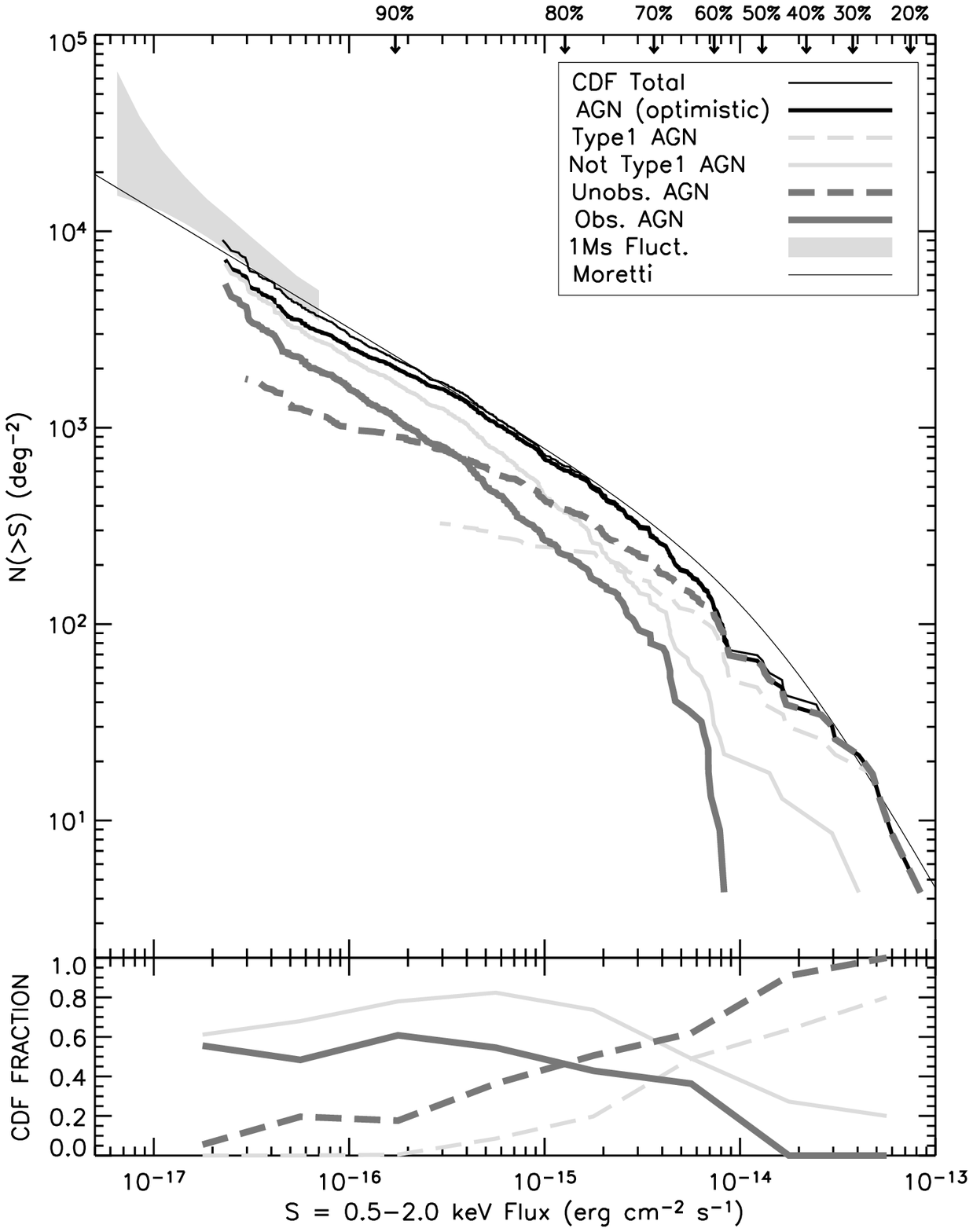}\hfill
\includegraphics[width=9.0cm]{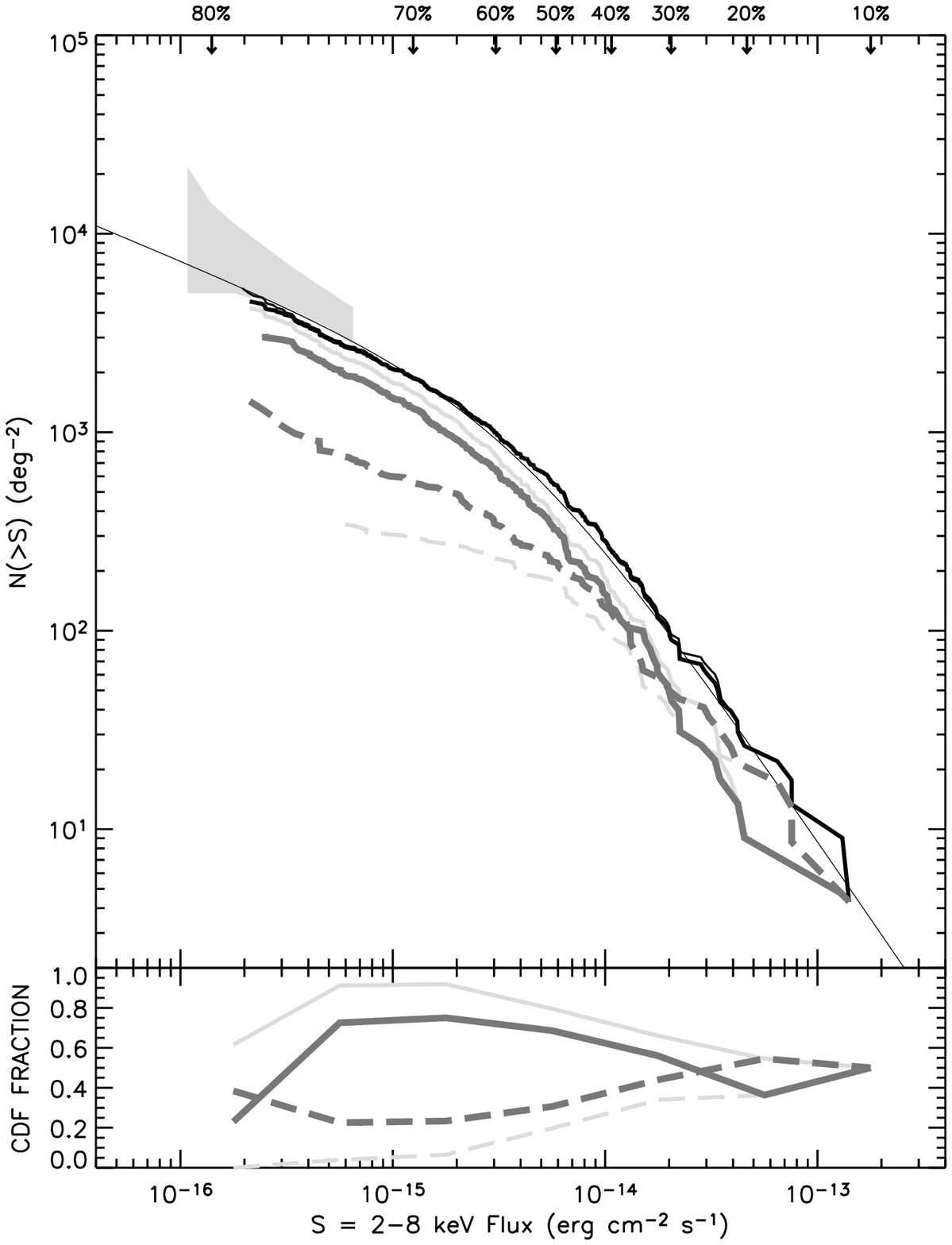}
}
\vspace{-0.1cm} \figcaption[Fig.8]{
  Plot identical to Fig.~\ref{fig:num_cnts_agn_gal} except we plot
  here only the number counts for the total (solid black curves) and
  subsets of our AGN class of objects: total AGN (thick solid black
  curves), optical Type~1 AGN (dashed light grey curves), optical
  ``not obviously Type~1'' AGN (solid light grey curves), X-ray
  unobscured/mildly obscured AGN (dashed dark grey curves), X-ray
  obscured AGN (solid dark grey curves).
\label{fig:num_cnts_agn}}
\vspace{-0.3cm}
\end{figure*} 

\subsubsection{AGN Number Counts}\label{sec:num_cnts_agn}

We further subdivided the AGN sample based on both optical and X-ray
properties. Unfortunately, the only meaningful optical division we can
place is based on AGN with broad lines reported in their optical
spectra (i.e., Type~1), and those apparently lacking them (i.e., not
obviously Type~1 AGN), since the optical spectra are not archived,
adequately analyzed, nor typically of very high signal-to-noise. Thus
the ``not obviously Type~1'' category is probably too loose to provide
much insight, as it is likely to include not only classic narrow-line
AGN (Type~2), but also slightly more luminous versions of
Compton-thick AGN like NGC~4945 and NGC~6240
\citep[e.g.,][]{Matt2000}, XBONGS \citep{Comastri2002}, and other odd
types. We further caution that spectral misclassifications are likely
for some sources since broad lines may be apparent in certain parts of
the optical spectrum but not others or may be missed due to
host-galaxy contamination or poor signal-to-noise data \citep[e.g.,
see discussion in][]{Moran2002}. In particular, there are a
significant number of CDF sources with high photometric redshifts that
could turn out to be broad-line AGN (see Figure~\ref{fig:R_vs_z}).
Given this likely incompleteness, the number counts for the optical
Type~1 and not obviously Type~1 AGN must be considered lower and upper
limits, respectively.  Similarly, AGN with column density estimates
$N_{\rm H}<10^{22}$~cm$^{-2}$ were considered X-ray unobscured/mildly
obscured AGN, and those above X-ray obscured AGN.  Again, we caution
that this distinction is far from exact, as it is based on the fit of
a relatively simple spectral model, which surely is inadequate
considering the known spectral complexity exhibited by local AGN
(e.g., reflection, scattering, partial covering, host-galaxy
contamination).  Such effects tend to contribute more soft flux than
predicted by a simple model and therefore can lead to an underestimate
of the column density or confuse the spectral fit.

From Figure~\ref{fig:num_cnts_agn}, we find that optical Type~1 AGN
fall off abruptly below $\sim$~$10^{-14}$~erg~cm$^{-2}$~s$^{-1}$ in
both bands, ultimately reaching source densities $\approx10$--20 times
lower than those of AGN that are not obviously Type~1 at the CDF
detection threshold. While this may be a legitimate effect, the
ability to detect broad lines tends to decrease at fainter optical
(and hence X-ray) fluxes, and thus we cannot exclude the possibility
that observational limitations contribute to this decline.
Considering AGN by their X-ray properties yields similar, albeit less
extreme, results. The source density of unobscured/mildly obscured AGN
gradually trails off below $\sim$~$10^{-14}$~erg~cm$^{-2}$~s$^{-1}$ in
both bands, ultimately reaching source densities $\approx2$--3 times
lower than those of obscured AGN at the CDF detection threshold.

The faint-end number-count slopes of all AGN subsamples are $\la1$. At
the CDF flux limits, we find AGN source densities of
7166$^{+304}_{-292}$ sources~deg$^{-2}$ (soft) and
4558$^{+216}_{-207}$ sources~deg$^{-2}$ (hard), which are factors of
$\sim10$--20 higher than the 300--500 sources deg$^{-2}$ found in the
deepest optical spectroscopic AGN surveys \citep[e.g.,][]{Hall2000,
  Steidel2002, Wolf2003, Hunt2004}. At face value, this means that
optical observations appear to be missing $>$90\% of AGN compared to
the deepest X-ray observations, although we concede that a direct
comparison is not entirely fair.  The optical surveys often target
specific redshift and luminosity ranges, and X-ray and optical survey
results are more consistent when compared within these ranges
\citep[c.f.,][]{Hunt2004}. Importantly though, the AGN discovered in
deep optical surveys are generally AGN with strong emission lines or
blue continua and only have source densities comparable to our optical
Type~1 AGN sample alone (i.e., $\approx$325--350 sources~deg$^{-2}$).
Given these facts, X-ray observations therefore appear to be much more
effective at identifying AGNs, particularly (more typical) obscured
AGNs. This conclusion is not entirely surprising given that the
optical spectra of many AGN in the CDFs are dominated by host-galaxy
light.

\subsubsection{AGN Completeness}\label{sec:agn_completeness}

While the AGN source densities achieved above are impressive, it is
important to understand the completeness of our X-ray selection of
AGN. It is likely, for instance, that we are still missing a
significant fraction of low-luminosity AGN (primarily those which have
steep X-ray spectra or fail to dominate over host-galaxy emission), as
such AGN are often difficult to identify even in the local Universe
\citep[e.g.,][]{Ho1997}. Local studies have also shown that $\sim$40\%
of moderate-to-high luminosity AGN are Compton thick, with very little
direct emission from the AGN \citep[e.g.,][]{Maiolino1998, Matt2000}.
Again, such sources are difficult to identify since their observed
X-ray luminosities are often $L_{\rm
  0.5-8.0~keV}<10^{42}$~erg~s$^{-1}$ and contaminated by host galaxy
emission; many such sources may be present in the CDFs but fall below
our AGN luminosity threshold.

Putting these caveats aside, however, there are only two classified
AGN in the CDFs which are not yet detected at X-ray wavelengths (both
in the 2~Ms \hbox{CDF-N}). The first, VLA~J123725.7$+$621128, is a
radio-bright ($\approx 6$ mJy at 1.4~GHz) wide-angle-tail source
estimated to lie between \hbox{$z\approx1$--2} with a rest-frame
0.5--2.0~keV luminosity limit of $\la5\times10^{41}$~erg~s$^{-1}$
\citep[e.g.,][]{Snellen2001,
  Bauer2002a}.\footnote{VLA~J123725.7$+$621128 can be marginally
  detected with some tweaking of WAVDETECT's detection parameters
  beyond the scope of A03.}  The second, 123720.0$+$621222, is a
narrow-line AGN at $z=2.445$ \citep{Hunt2004} with a rest-frame
0.5--2.0~keV luminosity limit of $\la2\times10^{42}$~erg~s$^{-1}$.

There are several more AGN candidates which lack X-ray detections,
although their classifications are considerably more tentative. Most
notably, 10 of the 30 radio sources with $\sim$1--10 mJy at 1.4~GHz
\citetext{\citealp{Richards2000}; A. Koekemoer 2004, private
  communication} lack X-ray detections in the combined CDFs.  This
number is consistent with expectations that $\sim$30\% of radio
sources at 1~mJy are star-forming galaxies
\citep[e.g.,][]{Jackson1999}, although four of these sources have
faint ($R>23$) optical counterparts (atypical of comparable
radio-detected star-forming galaxies) and thus may be radio-loud AGN.
At optical wavelengths, \citet{Wolf2004} report 51 ``QSO'' candidates
within the 1~Ms \hbox{CDF-S} region using the COMBO-17 photometric
redshift survey.  Nearly half (24) of these lack X-ray counterparts.
All but one of these non-detections, however, have ambiguous
classifications [i.e., ``QSO (Gal?)''] or $R>23.5$ (where the
reliability of the COMBO-17 photometric redshifts becomes poor). Thus,
we do not consider these to be convincing. Finally,
\citet{Sarajedini2003} discovered 16 (with a correction of $\pm2$ for
incompleteness and potential spurious sources) variable galactic
nuclei within the Hubble Deep Field-North.  These variable nuclei are
thought to be primarily low-luminosity AGN \citep[$-15\la M_{\rm
  B}\la-17$; e.g.,][]{Ho1997}, although a few may be nuclear
supernovae \citep[e.g.,][]{Riess1998}.  Only six have X-ray detections
in the 2~Ms \hbox{CDF-N}, indicating $\la$40\% overlap with the X-ray
sample.\footnote{ \citet{Sarajedini2003} actually report 7 X-ray
  matches, but the match of CXOHDFN J123651.73$+$621221.4 to
  HDF~3-461.9 is incorrect as this X-ray source lies outside of the
  galaxy and has a more viable faint near-IR counterpart. Also, the
  X-ray luminosities for several of the matches are low enough to be
  due entirely to star-formation rather than AGN activity.}  If all of
the variable nuclei are indeed AGN, then their number density
($\sim11000\pm2000$~deg$^{-2}$) would surpass that found in the CDFs.
It should be strongly emphasized here, however, that many of the AGN
selected via optical nuclear variability are likely to have X-ray
emission that is intrinsically 1--2 orders of magnitude lower than the
CDF detection limits \citep[i.e.,
$\sim10^{37}$--$10^{40}$~erg~s$^{-1}$; see][]{Ho2001}, and thus are
not directly comparable.

Given all of the above, X-ray emission appears to be one of the most
efficient and complete selection criteria currently available for
selecting moderate-to-high luminosity AGN.

\begin{figure*}
\vspace{-0.1in}
\centerline{
\includegraphics[width=9.0cm]{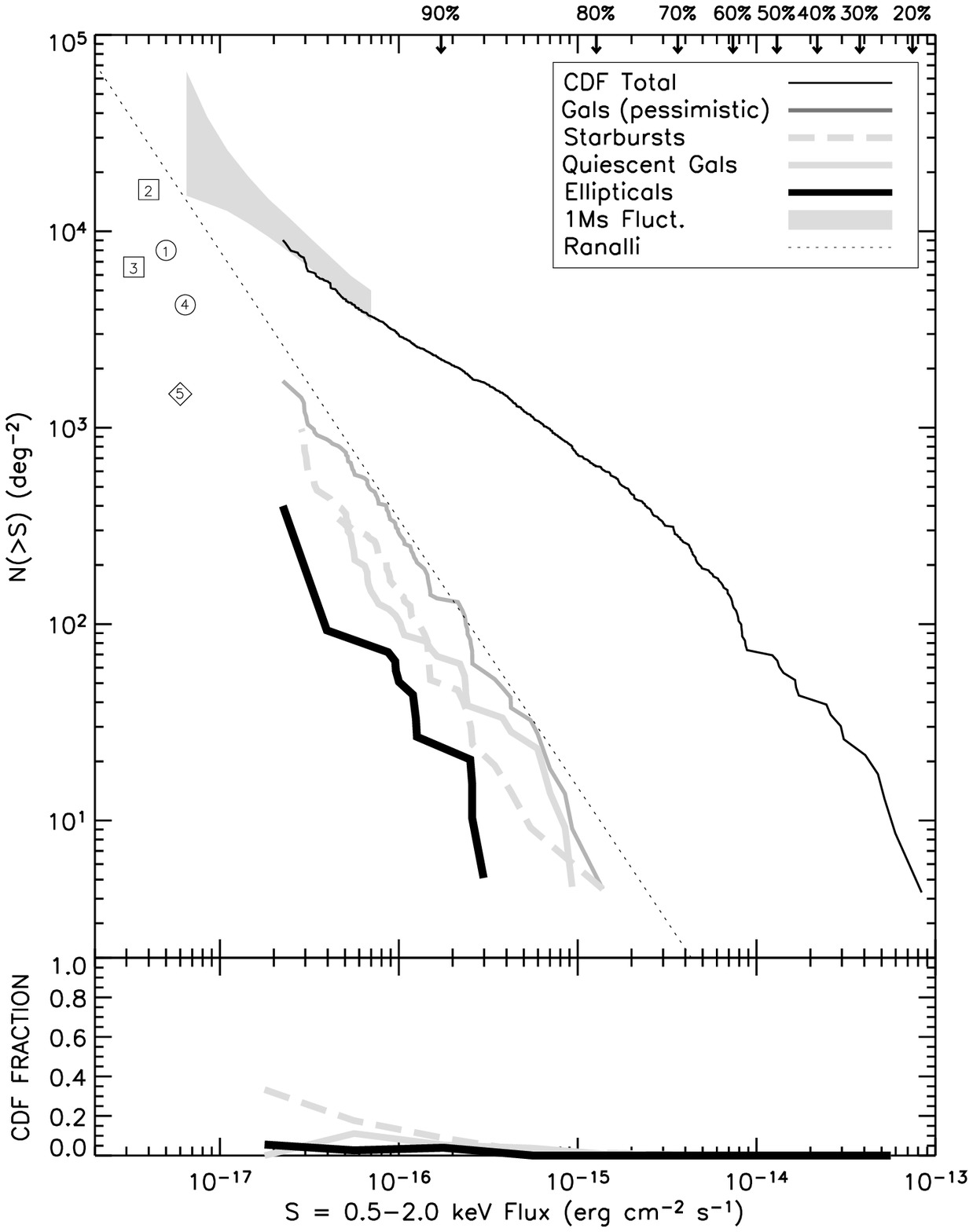}\hfill
\includegraphics[width=9.0cm]{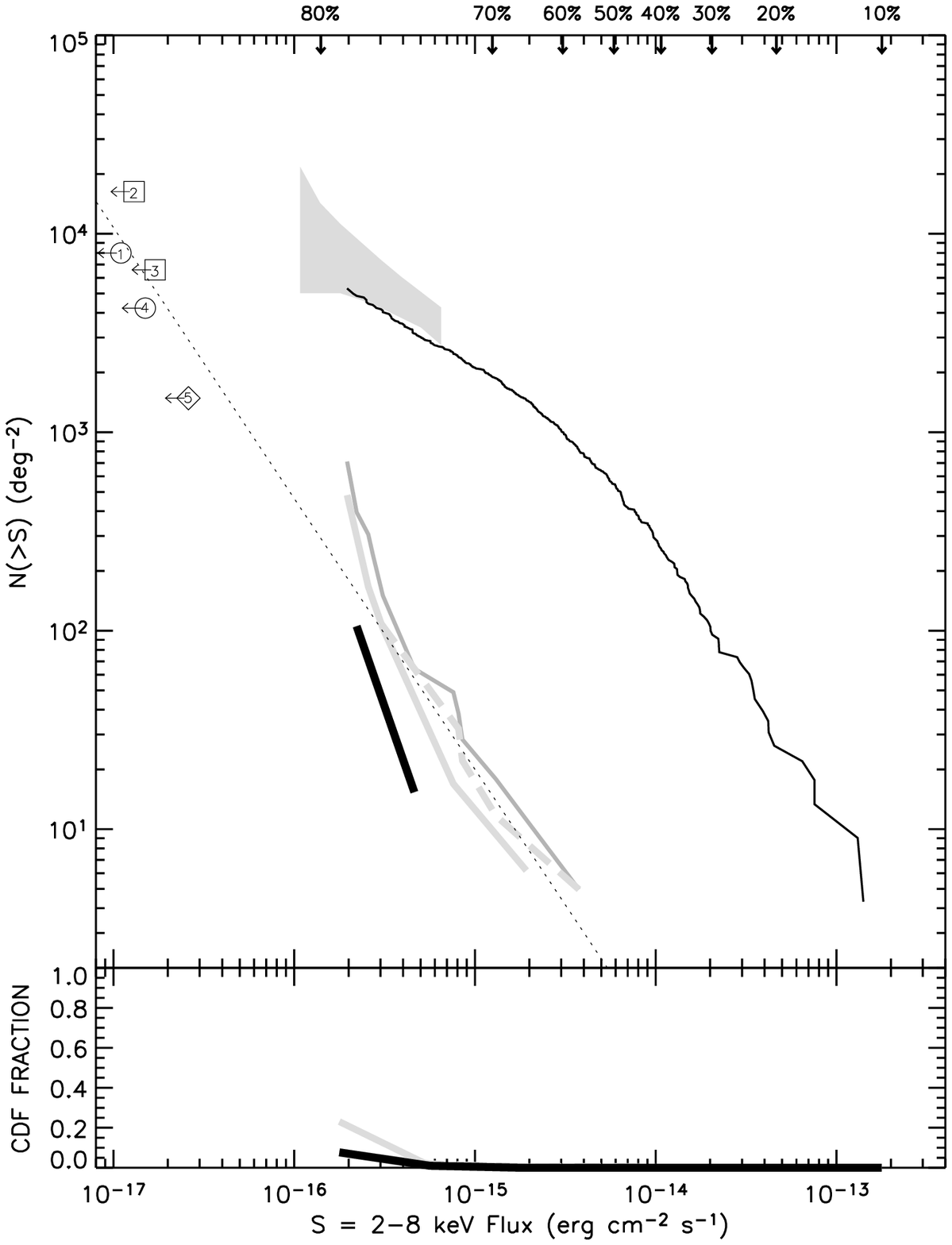}
}
\vspace{-0.1cm} \figcaption[Fig.9]{
  Plot identical to Fig.~\ref{fig:num_cnts_agn_gal} except we show
  here only the number counts for the total (solid black curves) and
  subsets of our galaxy class of objects: all galaxies (solid dark
  grey curves), starburst galaxies (dashed grey curves), quiescent
  galaxies (solid light grey curves), and elliptical galaxies (thick
  solid black curves).  Open symbols indicate the stacking results for
  (1) Lyman Break galaxies \citep{Brandt2001a}, (2) spiral galaxies
  \citep{Hornschemeier2002}, (3-4) Lyman and Balmer Break galaxies
  \citep{Nandra2002}, and (5) very red objects
  \citep[$I-K\ge4$;][]{Alexander2002a}.
\label{fig:num_cnts_gal}}
\vspace{-0.3cm}
\end{figure*} 

\subsubsection{Galaxy Number Counts}\label{sec:num_cnts_gals}

As noted above, the star-forming galaxy population is rising rapidly
below $\sim10^{-15}$~erg~cm$^{-2}$~s$^{-1}$ in both bands. As was done
for the AGN, we have further subdivided the galaxy sample based on
optical and X-ray properties to gain further insight into this
population.  We used the {\it HST}-derived elliptical galaxy sample of
\citet{Immler2004} to separate ellipticals (presumably passively
evolving at the redshifts of relevance here) from
spirals/irregulars/mergers (presumably actively star-forming), on the
assumption that the former is likely to be dominated by different
X-ray emission mechanisms (e.g., low-mass X-ray binaries,
low-luminosity AGN) compared to the latter (e.g., high-mass X-ray
binaries, supernovae, hot gas). The \citet{Immler2004} elliptical
galaxy sample is derived from the Great Observatories Origins Deep
Survey (GOODS) observations, which cover only $\approx$40\% of the
CDFs, so the ratio of ellipticals to other types is probably
underestimated. Fortunately, $82$\% of the galaxy sample overlaps with
the GOODS regions, so the elliptical galaxy number counts are likely
to be underestimated by $\la10$\%.  The non-ellipticals were split
into two groups above and below $f_{\rm 0.5-8.0~keV}/f_{\rm
  R}=0.01$, as set forth by \citet{Alexander2002b} to denote starburst
and quiescent galaxies, respectively.\footnote{We would prefer to use
  radio and infrared data, the "classical" probes of star-formation
  rate, to separate starburst and quiescent galaxies. However, the
  currently available infrared data cover only a tiny fraction of the
  \hbox{CDF-N} \citep[e.g.,][]{Alexander2002b}, and regions with
  overlapping 1.4~GHz and 8.5~GHz radio data \citep[necessary for
  radio spectral slopes; e.g.,][]{Bauer2002b} either cover only a
  small fraction of the field (\hbox{CDF-N}) or are not yet deep
  enough to detect large numbers of star-forming galaxies
  (\hbox{CDF-S}).}

The number-count slopes of the starburst and quiescent galaxy samples
are consistent with being Euclidean (see Table~\ref{tab:slopes}),
suggesting they may track to some extent the strongly evolving
infrared galaxy population \citep[e.g.,][]{Elbaz1999, Alexander2002b}.
The elliptical galaxies, meanwhile, have a somewhat flatter slope.
Caution should be exercised when interpreting the very steep slopes
observed in the hard band, of course, as these are uncertain and may be
biased upward by unidentified low-luminosity or Compton-thick AGNs in
all three samples. Quantifying the star-forming galaxy number counts
at brighter X-ray fluxes would be useful for comparison with our slope
estimates to determine potential evolutionary properties. We note
again that the starburst and quiescent galaxy number-count slopes
agree to within errors with the predictions made by
\citet{Ranalli2003} using both the X-ray/radio correlation
\citep[e.g.,][]{Bauer2002b, Ranalli2003} and the radio number counts
of \citet{Richards2000}.

Visual extrapolations of the starburst, quiescent, and elliptical
galaxy number counts to fainter X-ray fluxes appear to be consistent
with stacking results for various extragalactic source populations
(see Figure~\ref{fig:num_cnts_gal}). At the CDF flux limits, we find
source densities of 984$^{+151}_{-132}$, 343$^{ +68}_{ -57}$, and
105$^{+139}_{ -67}$ sources~deg$^{-2}$ (soft) and 123$^{ +84}_{ -53}$,
481$^{+327}_{-207}$, and 399$^{+144}_{-109}$ sources~deg$^{-2}$ (hard)
for the starburst, quiescent, and elliptical galaxy samples,
respectively.

\begin{figure*}
\vspace{-0.1in}
\centerline{
\includegraphics[width=9.0cm]{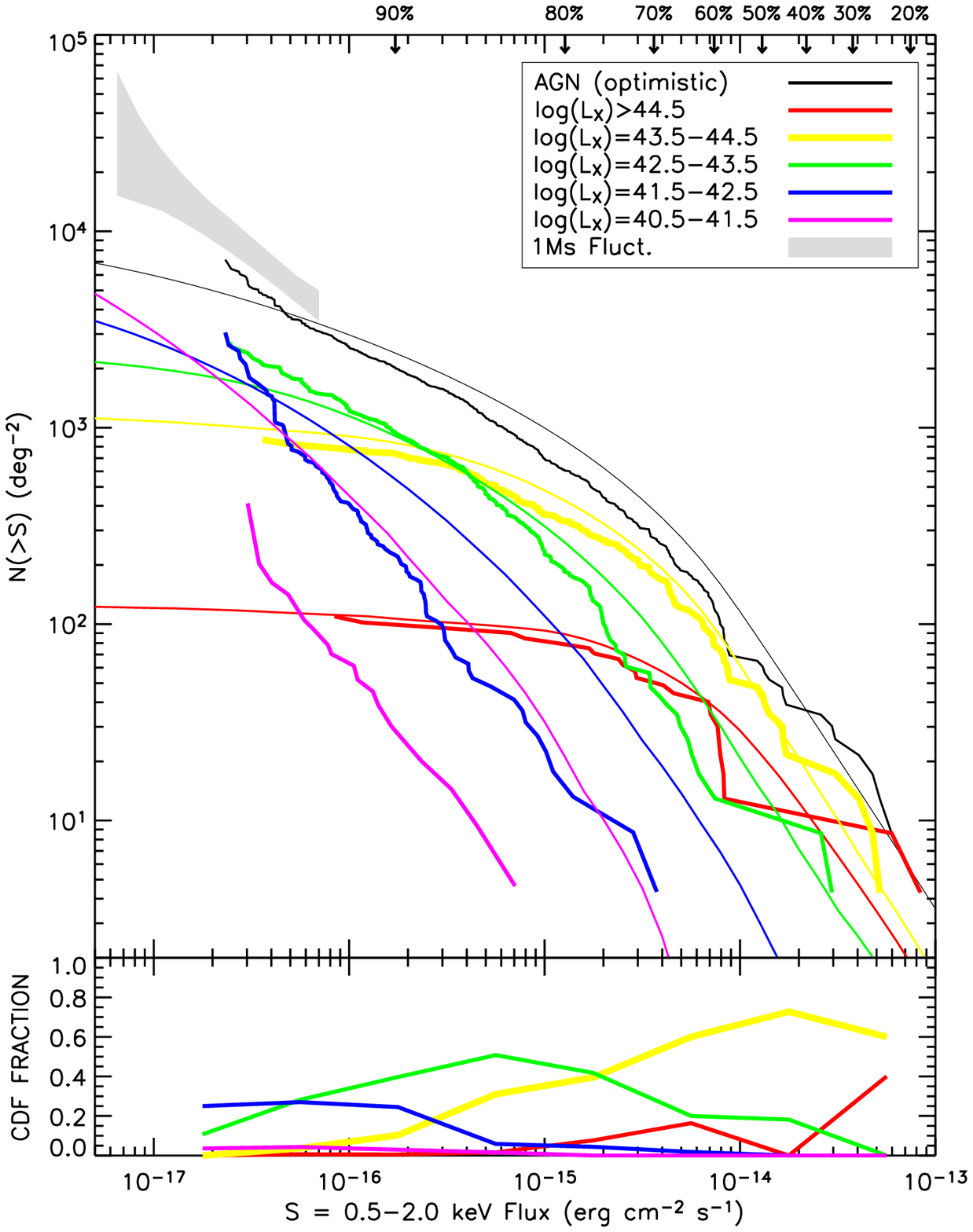}\hfill
\includegraphics[width=9.0cm]{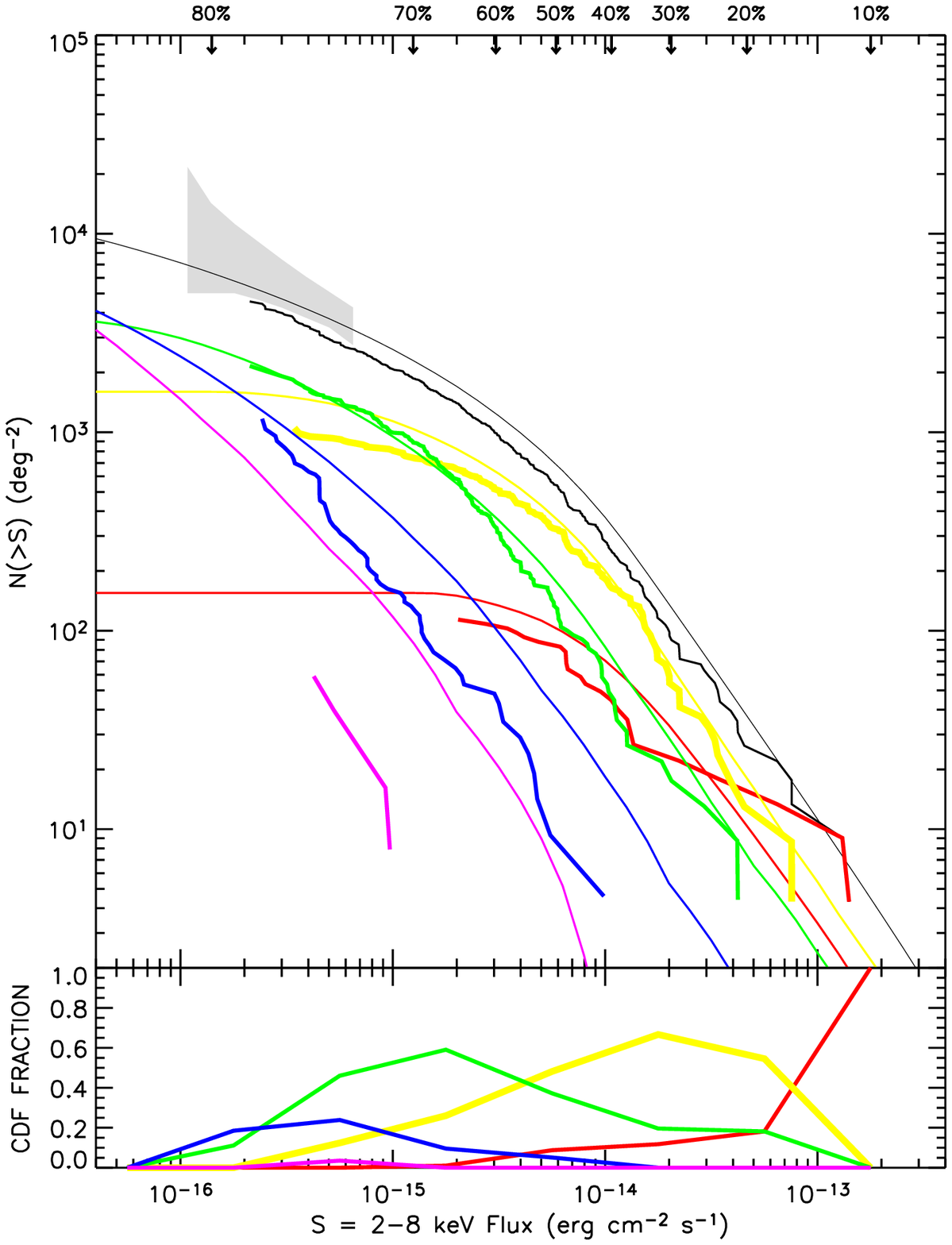}
}
\centerline{
\includegraphics[width=9.0cm]{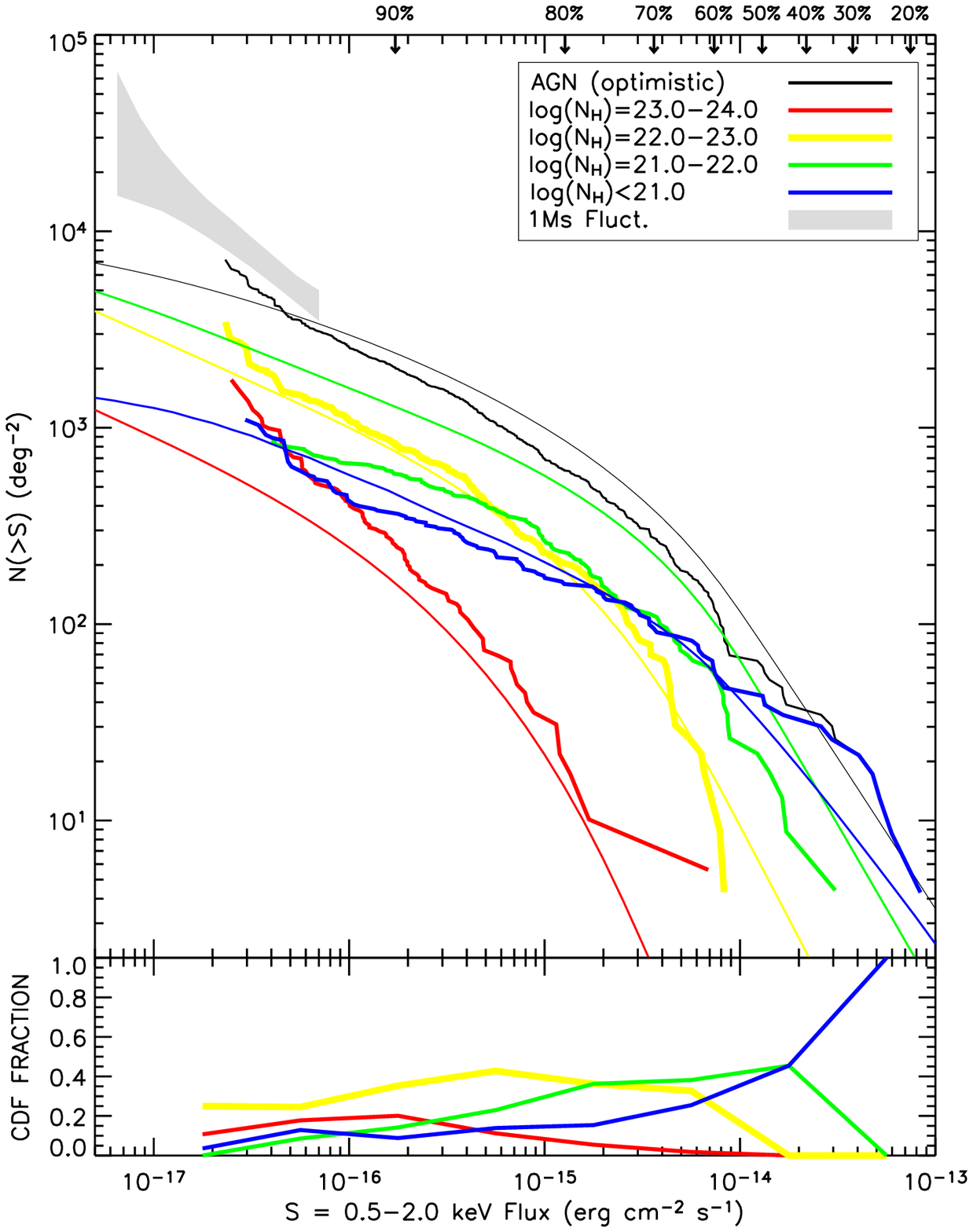}\hfill
\includegraphics[width=9.0cm]{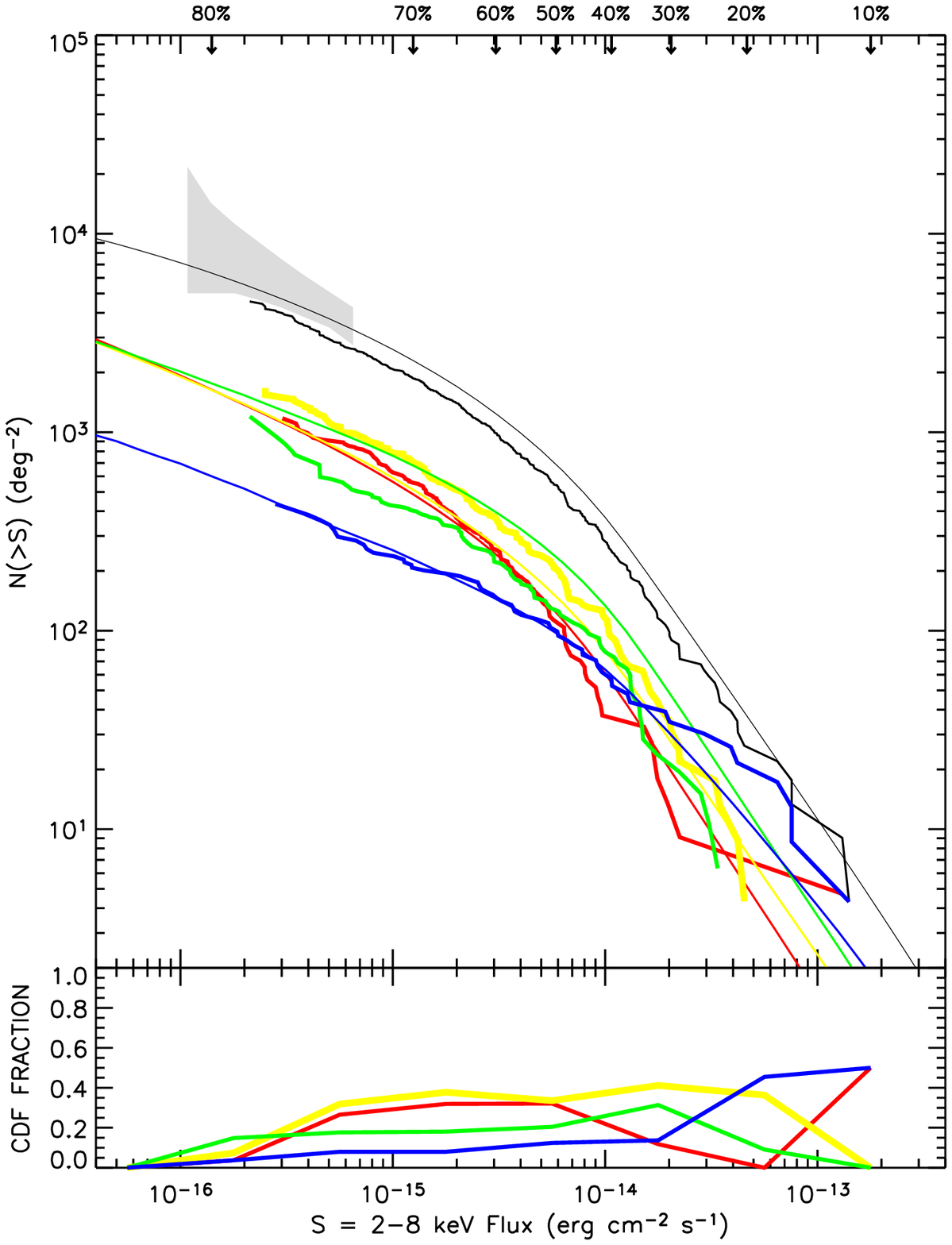}
}
\vspace{-0.1cm} 
\figcaption[Fig.10]{
  Plot identical to Fig.~\ref{fig:num_cnts_agn_gal} except we show
  here a comparison of the soft ({\it left}) and hard ({\it right})
  CDF number counts for all AGN (thick solid curves; see
  $\S$\ref{sec:num_cnts_agn}) and the predicted number counts from the
  \citet{Ueda2003} XLF (thin solid curves). The AGN number counts have
  been separated into several different intrinsic X-ray luminosity
  ({\it top}) and absorption ({\it bottom}) bins (color-coded
  following the order of the rainbow, with red denoting the highest
  luminosity and absorption bins) .
\label{fig:num_cnts_Lx}}
\vspace{-0.3cm}
\end{figure*} 

\subsection{AGN Number Counts by X-ray Luminosity and
  Absorption}\label{sec:lum_and_nh}

Using our estimated redshifts, we can also divide the CDF sources into
intrinsic X-ray luminosity and absorption bins. This is a useful
exercise both for better understanding the types of X-ray sources
contributing to the XRB and for testing the recent AGN X-ray
luminosity function of \citet[][hereafter simply the Ueda
XLF]{Ueda2003}. The Ueda XLF is constructed from X-ray sources with
fluxes above $F_{\rm
  2-8~keV}=3.2\times10^{-15}$~erg~cm$^{-2}$~s$^{-1}$ (including
sources from the \hbox{CDF-N}) and is strictly applicable only for
sources with intrinsic X-ray luminosities above $L_{\rm
  2-8~keV}=10^{41.5}$~erg~s$^{-1}$ and column densities below $N_{\rm
  H}=10^{24}$~cm$^{-2}$. To compare to our full CDF dataset, we have
extrapolated the Ueda XLF down to the CDF sensitivity limits and to
intrinsic luminosities of $L_{\rm 2-8~keV}=10^{40.5}$~erg~s$^{-1}$ in
order to test its predictive power. For details of this extrapolation, 
see \citet{Treister2004}.

Figure~\ref{fig:num_cnts_Lx} shows the optimistic AGN sample in the
CDFs, separated into several different intrinsic luminosity and
absorption classes. Also shown are the predicted X-ray number counts
from the Ueda XLF for the same classes. Some caution should be
exercised when interpreting the CDF number counts from $N_{\rm H}>
10^{23}$~cm$^{-2}$ AGN. {\it Chandra}'s effective area above 4--5~keV
declines rapidly such that the CDFs are almost certainly
systematically missing a significant fraction of the most highly
obscured AGN \citep[e.g.,][]{Treister2004}. This bias was not
accounted for in our simulations, as we averaged over the energy
dependence of {\it Chandra}'s effective area in our flux correction
and furthermore assumed that all of our simulated sources
had a spectral shape consistent with a $\Gamma=1.4$ power law.

It is readily apparent that the Ueda XLF overestimates the actual CDF
AGN number counts in both bands. We have derived error bars on the
cumulative distributions following \citet{Gehrels1986} and calculated
the deviation at each data point from the model in units of sigma. In
the soft band, the data points consistently lie below the model over
the flux range $2\times10^{-14}$~erg~cm$^{-2}$~s$^{-1}$ to
$5\times10^{-17}$~erg~cm$^{-2}$~s$^{-1}$, with the statistical
deviation increasing from 1--2$\sigma$ at the brighter end of this
range to 4--5$\sigma$ at the fainter end of this range. In the hard
band, the data points consistently lie below the model over the entire
flux range, with the statistical deviation steadily increasing with
decreasing flux from 1--5.5$\sigma$. The discrepancies are still
significant (up to 3$\sigma$) even if we only consider sources over
the original flux range used to construct the Ueda XLF. Note that the
upturn seen in the soft-band data below $F_{\rm
  0.5-2.0~keV}\sim5\times10^{-17}$~erg~cm$^{-2}$~s$^{-1}$ in
Figure~\ref{fig:num_cnts_Lx} is likely due to X-ray spectral
complexity (as discussed in $\S$\ref{sec:classification}) not
accounted for in the simple spectral model used to construct the Ueda
XLF.

Turning to the individual intrinsic luminosity bins in
Figure~\ref{fig:num_cnts_Lx}, we find that the soft-band XRB is
dominated by sources with $L_{\rm 0.5-8~keV}>10^{43.5}$~erg~s$^{-1}$,
while the hard-band XRB is dominated by somewhat less luminous sources
with $L_{\rm 0.5-8~keV}=10^{42.5}$--$10^{44.5}$~erg~s$^{-1}$.  There
is good agreement between the highest individual CDF luminosity bins and
the Ueda XLF in both bands, with the best estimates in the soft band
being among the $L_{\rm 0.5-8~keV}>10^{44.5}$~erg~s$^{-1}$ and $L_{\rm
  0.5-8~keV}=10^{42.5}$--$10^{43.5}$~erg~s$^{-1}$ AGN samples, as well
as the bright end of the $L_{\rm
  0.5-8~keV}=10^{43.5}$--$10^{44.5}$~erg~s$^{-1}$ AGN sample. The
Ueda XLF, however, consistently overestimates the soft-band number
counts of (1) $L_{\rm 0.5-8~keV}=10^{43.5}$--$10^{44.5}$~erg~s$^{-1}$
AGN below $1\times10^{-14}$~erg~cm$^{-2}$~s$^{-1}$, with the
statistical deviation steadily increasing with decreasing flux from
1--6$\sigma$, and (2) $L_{\rm 0.5-8~keV}<10^{42.5}$~erg~s$^{-1}$ AGN
over the flux range $2\times10^{-14}$~erg~cm$^{-2}$~s$^{-1}$ to
$5\times10^{-17}$~erg~cm$^{-2}$~s$^{-1}$, with the statistical
deviation increasing from 1--2$\sigma$ at the brightest and faintest
ends of this range to 8--10$\sigma$ in the middle of this range.  In
the hard band, the Ueda XLF again consistently overestimates the
number counts of the (1) $L_{\rm
  0.5-8~keV}=10^{43.5}$--$10^{44.5}$~erg~s$^{-1}$ AGN below
$1\times10^{-14}$~erg~cm$^{-2}$~s$^{-1}$, with the statistical
deviation steadily increasing with decreasing flux from 1--6$\sigma$,
and (2) $L_{\rm 0.5-8~keV}<10^{42.5}$~erg~s$^{-1}$ AGN over the flux
range $2\times10^{-14}$~erg~cm$^{-2}$~s$^{-1}$ to
$5\times10^{-17}$~erg~cm$^{-2}$~s$^{-1}$, with the statistical
deviation increasing from 1--2$\sigma$ at the brightest and faintest
ends of this range to 5--6$\sigma$ in the middle of this range.  The
discrepancies between the Ueda XLF predictions and the actual
number of low-luminosity AGN have also been noted by
\citet{Menci2004}.

In terms of intrinsic absorption bins, we find that in the soft band
the XRB is dominated by relatively unobscured sources with $N_{\rm
  H}<10^{22}$~cm$^{-2}$, while in the hard band the contribution to
the XRB from all of the different absorption bins is nearly equal.
Again, there is good agreement between some of the individual CDF
absorption bins and the Ueda XLF, although discrepancies between the
Ueda XLF and the data are more evident here. In the soft band, the
agreement is good for $N_{\rm H}<10^{21}$~cm$^{-2}$ and $N_{\rm
  H}=10^{22}$--$10^{23}$~cm$^{-2}$ AGN. However, the Ueda XLF model
consistently overestimates the soft-band number counts of $N_{\rm
  H}=10^{21}$--$10^{22}$~cm$^{-2}$ AGN over the entire flux range,
with the statistical deviation steadily increasing with decreasing
flux from 1--19$\sigma$, and underestimates the soft-band number
counts of $N_{\rm H}=10^{23}$--$10^{24}$~cm$^{-2}$ AGN over the entire
flux range, with the statistical deviation steadily increasing with
decreasing flux from 1--6$\sigma$. Note that this latter discrepancy
is in the opposite sense to {\it Chandra}'s possible observational
bias mentioned above and is therefore likely to be even worse. In the
hard band, the agreement between the Ueda XLF model and the data is
good for $N_{\rm H}<10^{21}$~cm$^{-2}$, $N_{\rm
  H}=10^{22}$--$10^{23}$~cm$^{-2}$, and $N_{\rm
  H}=10^{23}$--$10^{24}$~cm$^{-2}$ AGN, but consistently overestimates
the hard-band number counts of $N_{\rm
  H}=10^{21}$--$10^{22}$~cm$^{-2}$ AGN over the flux range
$2\times10^{-14}$~erg~cm$^{-2}$~s$^{-1}$ to
$4\times10^{-16}$~erg~cm$^{-2}$~s$^{-1}$, with the statistical
deviation steadily increasing with decreasing flux from 1--6$\sigma$.

\begin{figure}
\vspace{-0.1in}
\centerline{
\includegraphics[width=9.0cm]{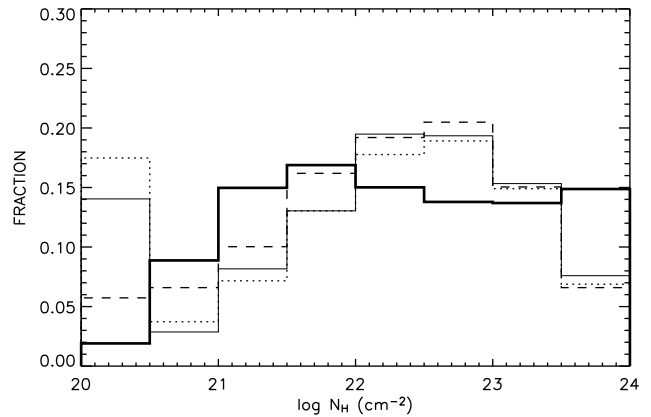}
}
\vspace{-0.1cm} 
\figcaption[Fig.11]{
  We show a comparison of the $N_{\rm H}$ distribution from spectral
  analysis of the CDFs (thin solid histogram) versus that predicted
  from the \citet{Ueda2003} XLF (thick solid histogram). Also shown
  are the 1~$\sigma$ upper and lower limits of the CDF $N_{\rm H}$
  distribution (dashed and dotted histograms, respectively),
  indicating that the shape of the CDF distribution is relatively
  robust except for the lowest $N_{\rm H}$ bin. We caution that the
  actual $N_{\rm H}$ distribution may be skewed towards higher $N_{\rm
    H}$ values if the redshift distribution we have assumed for the
  sources lacking redshifts is higher or our spectral models are too
  simplistic (i.e., they fail to account for known spectral complexity
  in a systematic way). In the former case the shift should not be
  more than a factor of a few, while in the latter such spectral
  complexity would apply to both the CDF and Ueda distributions.
\label{fig:cdf_nh}}
\vspace{-0.3cm}
\end{figure} 

Some of the discrepancies between the individual luminosity and
absorption bins and the Ueda XLF may be due to our assumptions
regarding the redshift distribution (and hence the absorption and
X-ray luminosity distributions).  Several arguments suggest that this
is unlikely to be the case: (1) The Ueda XLF overestimates the total
number counts as well, where no redshift information is used.  (2) The
agreement between the lowest and highest column density sources is
fairly good. Significant changes to the redshift distribution either
way would worsen the agreement.  (3) If the redshift distribution is
on average higher than we have estimated in
$\S$\ref{sec:num_cnts_type}, then sources from lower X-ray luminosity
bins, where the overestimate is already bad, would move into higher
ones where the agreement is already acceptably good (thus worsening
the agreement).  (4) The X-ray sources are unlikely to lie at
significantly lower redshifts than those assumed given that the
sources are not generally detected in the bluest optical bands,
implying substantial redshifts or that they reside in severely
underluminous galaxies. The latter is unlikely given what we know
about the host galaxies of powerful AGN in the nearby Universe
\citep{Kauffmann2003}.

Figure~\ref{fig:cdf_nh} shows the derived CDF $N_{\rm H}$ distribution
(including errors) and the extrapolated \citet{Ueda2003} model. There
is rough agreement between the data and model, although the CDF
distribution clearly appears to be bimodel, while the \citet{Ueda2003}
model does not. These deviations are less obvious in
Figure~\ref{fig:num_cnts_Lx} because of the coarser $N_{\rm H}$
binning and the effect contamination from star-forming host galaxies
has on the X-ray fluxes of highly obscured AGN. While the upper and
lower limits on the overall $N_{\rm H}$ determinations (dashed and
dotted lines, respectively) are tight enough to exclude the
possibility of this bimodality being spurious, the assumed redshift
distribution for sources lacking redshifts could play a role in
shaping the $N_{\rm H}$ distribution. For instance, adopting a larger
scatter in the redshift distribution might broaden the $N_{\rm H}$
distribution to be more in line with the flat distribution from the
\citet{Ueda2003} model. Alternatively, if we have systematically
underestimated redshifts this would tend to shift the second $N_{\rm
  H}$ peak toward higher values and increase discrepancies between the
data and model at moderate column densities. We consider the CDF
$N_{\rm H}$ distribution shown in Figure~\ref{fig:cdf_nh} to be fairly
robust, however, since it would take exceptionally large deviations in
the adopted redshift distribution to change intrinsic $N_{\rm H}$
values by more than a factor of a few given that $N_{\rm
  H}(intr)\approx(1+z)^2.7 N_{\rm H}(obs)$. Thus, it appears that the
extrapolated \citet{Ueda2003} $N_{\rm H}$ distribution is not entirely
appropriate for the CDF data. \citet{Treister2004} come to a similar
conclusion using a more luminous subset of the CDF AGN and an $N_{\rm
  H}$ distribution derived instead from X-ray colors.

\subsection{Contributions to the Extragalactic X-ray 
Background}\label{sec:contrib_xrb}

The contributions to the soft and hard XRB from each CDF source type
are shown in Table~\ref{tab:stats}, normalized by the average total
flux density of the XRB [taken from M03 in the soft band to be
($7.52\pm0.35)\times10^{-12}$~erg~cm$^{-2}$~s$^{-1}$~deg$^{-2}$ and
from \citealp{DeLuca2004} in the hard band to be
($2.24\pm0.11)\times10^{-11}$~erg~cm$^{-2}$~s$^{-1}$~deg$^{-2}$].  For
sources brighter than observed in the CDFs (i.e., $F_{\rm
  0.5-2.0~keV}>8\times10^{-14}$~erg~cm$^{-2}$~s$^{-1}$ and $F_{\rm
  2-8~keV}>1\times10^{-13}$~erg~cm$^{-2}$~s$^{-1}$), we have adopted
total resolved flux-density contributions from the models of M03,
which are 19.2\% in the soft band and 11.5\% in the hard band; we
assume uncertainties of $\approx$4\% on these values estimated from
Figure~5 of M03. 

The crude breakdown of these bright resolved fractions is 40\% stars,
40\% Type~1 AGN, 5\% Type~2 AGN, 10\% clusters, 5\% galaxies in the
soft band, and 1\% stars, 69\% Type~1 AGN, 20\% Type~2 AGN, 8\%
clusters, and 2\% galaxies in the hard band
\citep[e.g.,][]{Krautter1999, LaFranca2002, Akiyama2003}.  When known,
these values have been added to column~4 of Table~\ref{tab:stats} to
produce total resolved XRB fractions in column~5 of
Table~\ref{tab:stats}. In total, and neglecting possible large-scale
structure effects and instrumental cross-calibration uncertainties, we
find 89.5$^{+5.9}_{-5.7}$\% and 86.9$^{+6.6}_{-6.3}$\% of the
XRB has been resolved in the soft and hard bands, respectively.

We find that AGN as a whole contribute $\approx$83\% and $\approx$95\%
to the resolved soft and hard XRBs, respectively. Dividing the AGN
further, we find that Type~1 AGN alone contribute significantly
($\sim57$\%) to the AGN fraction of the resolved XRB in the soft band,
but far less so ($\sim35$\%) in the hard band.  X-ray
unobscured/mildly obscured AGN completely dominate ($\sim76$\%) the
AGN fraction of the resolved XRB in the soft band, and contribute
nearly half ($\sim46$\%) in the hard band.  If this latter trend
continues to harder energies, it suggests that highly obscured X-ray
sources will likely dominate the XRB emission at higher energies
(e.g., closer to $\approx20$--40~keV peak of the XRB).  This can be
directly tested by performing similar analyses on deep 5.0--10.0~keV
number counts from {\it XMM-Newton}. In contrast, star-forming
galaxies comprise only $\approx$10\% and $\approx$2\% of the soft and
hard XRBs. Starbursts appear to make up the bulk of the X-ray emission
in the soft band, while quiescent galaxies apparently dominate in the
hard band; contamination from embedded, highly obscured AGN may be
responsible in the latter case.

Extrapolating the total number-count slopes from
Table~\ref{tab:slopes} down to
$1\times10^{-18}$~erg~cm$^{-2}$~s$^{-1}$ in the soft band and
$1\times10^{-17}$~erg~cm$^{-2}$~s$^{-1}$ in the hard band can account
for an additional $6.6^{ +8.7}_{ -6.3}$\% and $24.1^{+38.8}_{-13.8}$\%
of the XRBs, respectively. We note, however, that some of the
number-count slopes for the individual CDF source types are
significantly steeper than the overall number-count slopes and that
independent extrapolation of these slopes actually exceeds the total
XRB by a large percentage in some cases. This suggests that (1) known
point-source populations can explain all of the remaining unresolved
XRB and (2) the steep slopes found for some source types must flatten
in the decade of flux below the CDF sensitivity limits. As such, we
contend that there is little room for extra contributions from, e.g.,
a truly diffuse component. We additionally find that the number
density of star-forming galaxies should overtake that of the AGN just
below $F_{\rm 0.5-2.0~keV}\sim1\times10^{-17}$~erg~cm$^{-2}$~s$^{-1}$
in both bands, and these sources should contribute an additional
$\sim$10\% and $\sim$5\% to the soft and hard XRB emission \citep[here
we have assumed a flattening of the starburst number counts for
sources as inferred from infrared galaxy number counts;
e.g.,][]{Chary2001}.

Thus, the overall contribution of galaxies to the XRB appears to be
10--20\% in the soft band and 2--6\% in the hard band, indicating that
if the cosmic star-formation rate evolves as $(1+z)^{q}$ then $q$ is
likely to be near $q=3$ rather than a higher value
\citep[see][]{Persic2003}.  This is consistent with previous results
from stacking analyses \citep[e.g.,][]{Hornschemeier2002,
  Georgakakis2004} as well as the recently measured normal galaxy XLF
\citep{Norman2004}.

We note that we are using relatively broad X-ray bands, and that the
resolved fraction at a given energy may change somewhat throughout
these bandpasses. Since the effective area of {\it Chandra} falls off
rapidly above 4~keV, we are likely to resolve more of the 2--4~keV XRB
and less of the 4--8~keV XRB.  An investigation into this is beyond
the scope of this study, but has been undertaken by \citet[2004;][in
preparation]{Worsley2004} using the deep Lockman Hole and the CDFs.
They perform photometry of the resolved source in several narrow X-ray
bands, finding that the resolved fraction of the XRB is indeed
sytematically lower at harder X-ray energies. They also find that the
resultant X-ray spectrum of the unresolved XRB is consistent with that
from a highly-obscured ($N_{\rm H}\sim10^{23}$--$10^{24}$~cm$^{-2}$
for $z<2$) AGN.

\section{Conclusions}\label{sec:conclusion}

We have combined the \hbox{CDF-N} and \hbox{CDF-S} X-ray samples, along with
substantial public ancillary data, to examine the X-ray number counts
as a function of source type. Extensive simulations were carried out
to quantify and correct for the completeness and flux bias problems
which affect the CDF number counts. Once corrected, we find that the
number counts from the two fields are consistent with each other
aside from sources detected in the 2--8~keV band below $F_{\rm
  2-8~keV}\approx1\times10^{-15}$~erg~cm$^{-2}$~s$^{-1}$, where
statistical deviations gradually increase to $3.9\sigma$ at the
faintest flux levels. We also find that our overall number counts are
consistent with previous determinations.  In total, we have resolved
89.5$^{+5.9}_{-5.7}$\% and 86.9$^{+6.6}_{-6.3}$\% of the
extragalactic 0.5--2.0~keV and 2--8~keV XRB, respectively.

Using a classification scheme based on X-ray spectral properties,
intrinsic X-ray luminosities, radio morphologies, variability, optical
spectroscopic classifications, and X-ray-to-optical flux ratios, we
have separated the CDF X-ray sources into 698 AGN (split into optical
Type~1 and not obviously Type~1, and X-ray obscured and
unobscured/mildly obscured), 109 star-forming galaxies (split into
starburst, quiescent, and elliptical), and 22 Galactic stars, and
determined their individual number counts.  We additionally calculate
the number-count slopes and normalizations below $2\times10^{-15}$
erg~cm$^{-2}$~s$^{-1}$ for all source types assuming a single
power-law model.

We confirm that AGN power the bulk of the extragalactic XRB, with only
a small contribution from star-forming galaxies.  The most significant
contributions to the XRB are from sources with $L_{\rm
  0.5-8~keV}>10^{43.5}$~erg~s$^{-1}$ and $N_{\rm H}<10^{22}$~cm$^{-2}$
in the soft band and $L_{\rm
  0.5-8~keV}=10^{42.5}$--$10^{44.5}$~erg~s$^{-1}$ and an
evenly-distributed range of absorption column densities in the hard
band. This trend suggests that even less luminous, more highly
obscured AGN may in fact dominate the number counts at higher
energies, where the XRB intensity peaks. At the CDF flux limits, the
overall AGN source densities are 7166$^{+304}_{-292}$
sources~deg$^{-2}$ and 4558$^{+216}_{-207}$ sources~deg$^{-2}$,
respectively, factors of $\sim10$--20 higher than found in the deepest
optical spectroscopic surveys.

While star-forming galaxies make up a small fraction of sources with
fluxes higher than $\sim10^{-15}$~erg~cm$^{-2}$~s$^{-1}$ in both
bands, their numbers climb steeply below this flux such that they
eventually achieve source densities of 1727$^{+187}_{-169}$
sources~deg$^{-2}$ and 711$^{+270}_{-202}$ sources~deg$^{-2}$ at the
CDF flux limits (with starburst galaxies making the largest
contribution) and comprise up to $\sim40$\% of the sources at the
faintest X-ray fluxes.  Extrapolation of the number-count slopes for
galaxy source types can account for all of the remaining unresolved
soft and hard XRBs.  Moreover, within a factor of a few below the
current CDF soft-band flux limit, the sky density of star-forming
galaxies will likely overtake that of AGN.

We have also compared the \citet{Ueda2003} XLF with our X-ray number
counts. While the agreement is generally good for most subsets of CDF
sources, it appears that the extrapolated XLF requires some
significant refinements before it will be able to reproduce the data
with sufficient accuracy.

Further improvements to our study can be made in several ways.
Constraining the bright-flux end of the number counts for all of our
source types would be particularly useful for determining more
accurate slopes and XRB contributions. Additional observations of the
2~Ms \hbox{CDF-N} with {\it Chandra} would (1) yield additional
photons to discriminate better between AGN and galaxies and model
spectral complexities for better absorption column density and
luminosity estimates, (2) allow confirmation of the sharp rise in the
number counts of both star-forming galaxies and moderate-to-high
obscuration ($N_{\rm H}\sim10^{22}-10^{24}$) AGN that we find here,
and (3) perhaps provide evidence for their eventual flattening as is
required by the overall XRB flux density. Upcoming ancillary
observations at infrared ({\it Spitzer}) and radio (VLA) wavelengths,
as well as detailed analyses of the existing optical spectra, would be
extremely helpful for refining our AGN and galaxy subclassifications.
Finally, a high-sensitivity, high spatial resolution X-ray telescope
able to probe to $10^{-15}$~erg~cm$^{-2}$~s$^{-1}$ in the 10--40~keV
band is desperately needed to resolve the peak of the XRB and
determine once and for all the nature and composition of the XRB.

\acknowledgements This work would not have been possible without the
support of the entire {\it Chandra} and ACIS teams; we particularly
thank P.~Broos and L.~Townsley for data analysis software and CTI
correction support.
We thank S. Allen, A. Fabian, and W. Yuan for useful discussions.
We acknowledge the financial support of NSF CAREER award AST-9983783
(FEB, WNB), CXC grant GO2-3187A (FEB, WNB), PPARC (FEB), 
the Royal Society (DMA), 
NSF grant AST03-07582 (DPS), 
the Fundaci\'on Andes (ET),
Chandra fellowship grant PF2-30021 (AEH), 
and NASA grants NAS~8-38252 and NAS~8-01128 (GPG, PI).


\end{document}